\documentclass[twocolumn]{aastex63}
\usepackage{xcolor}

\usepackage{natbib} 
\usepackage{amsmath}
\usepackage{enumitem}
\shorttitle{Atmospheric astrometric distortions}
\shortauthors{Fortino et al.}
\reportnum{FERMILAB-PUB-20-555-AE}

\def\eqq#1{Equation~(\ref{#1})}

\def\ie{{\it i.e.}}
\def\eg{{\it e.g.}}

\newcommand{\vpi}{\ensuremath{\boldsymbol{\pi}}}
\newcommand{\vx}{\ensuremath{\mathbf{x}}}
\newcommand{\vu}{\ensuremath{\mathbf{u}}}
\newcommand{\vU}{\ensuremath{\mathbf{U}}}
\newcommand{\vn}{\ensuremath{\mathbf{n}}}
\newcommand{\vk}{\ensuremath{\mathbf{k}}}
\newcommand{\vs}{\ensuremath{\mathbf{s}}}
\newcommand{\matK}{\ensuremath{\mathbf{K}}}
\newcommand{\matM}{\ensuremath{\mathbf{M}}}
\newcommand{\matC}{\ensuremath{\mathbf{C}}}

\newcommand{\des}{\textit{DES}}
\newcommand{\lsst}{\textit{LSST}}
\newcommand{\gaia}{\textit{Gaia}}

\begin{document}

\reportnum{arXiv submission, 26 Oct 2020}

\title{Reducing ground-based astrometric errors with \gaia\ and Gaussian processes}

\suppressAffiliations

\author[0000-0001-7559-7890]{W.~F.~Fortino}
\affiliation{Department of Physics and Astronomy, University of
  Pennsylvania, Philadelphia, PA 19104, USA}
\affiliation{Department of Physics and Astronomy,
University of Delaware,
Newark, DE 19716, USA}
\email{fortino@sas.upenn.edu}

\author[0000-0002-8613-8259]{G.~M.~Bernstein}
\affiliation{Department of Physics and Astronomy, University of Pennsylvania, Philadelphia, PA 19104, USA}
\email{garyb@physics.upenn.edu}

\author[0000-0003-0743-9422]{P.~H.~Bernardinelli}
\affiliation{Department of Physics and Astronomy, University of Pennsylvania, Philadelphia, PA 19104, USA}


\author{M.~Aguena}
\affiliation{Departamento de F\'isica Matem\'atica, Instituto de F\'isica, Universidade de S\~ao Paulo, CP 66318, S\~ao Paulo, SP, 05314-970, Brazil}
\affiliation{Laborat\'orio Interinstitucional de e-Astronomia - LIneA, Rua Gal. Jos\'e Cristino 77, Rio de Janeiro, RJ - 20921-400, Brazil}
\author{S.~Allam}
\affiliation{Fermi National Accelerator Laboratory, P. O. Box 500, Batavia, IL 60510, USA}
\author{J.~Annis}
\affiliation{Fermi National Accelerator Laboratory, P. O. Box 500, Batavia, IL 60510, USA}
\author{D.~Bacon}
\affiliation{Institute of Cosmology and Gravitation, University of Portsmouth, Portsmouth, PO1 3FX, UK}
\author{K.~Bechtol}
\affiliation{Physics Department, 2320 Chamberlin Hall, University of Wisconsin-Madison, 1150 University Avenue Madison, WI  53706-1390}
\author{S.~Bhargava}
\affiliation{Department of Physics and Astronomy, Pevensey Building, University of Sussex, Brighton, BN1 9QH, UK}
\author{D.~Brooks}
\affiliation{Department of Physics \& Astronomy, University College London, Gower Street, London, WC1E 6BT, UK}
\author{D.~L.~Burke}
\affiliation{Kavli Institute for Particle Astrophysics \& Cosmology, P. O. Box 2450, Stanford University, Stanford, CA 94305, USA}
\affiliation{SLAC National Accelerator Laboratory, Menlo Park, CA 94025, USA}
\author{J.~Carretero}
\affiliation{Institut de F\'{\i}sica d'Altes Energies (IFAE), The Barcelona Institute of Science and Technology, Campus UAB, 08193 Bellaterra (Barcelona) Spain}
\author{A.~Choi}
\affiliation{Center for Cosmology and Astro-Particle Physics, The Ohio State University, Columbus, OH 43210, USA}
\author{M.~Costanzi}
\affiliation{INAF-Osservatorio Astronomico di Trieste, via G. B. Tiepolo 11, I-34143 Trieste, Italy}
\affiliation{Institute for Fundamental Physics of the Universe, Via Beirut 2, 34014 Trieste, Italy}
\author{L.~N.~da Costa}
\affiliation{Laborat\'orio Interinstitucional de e-Astronomia - LIneA, Rua Gal. Jos\'e Cristino 77, Rio de Janeiro, RJ - 20921-400, Brazil}
\affiliation{Observat\'orio Nacional, Rua Gal. Jos\'e Cristino 77, Rio de Janeiro, RJ - 20921-400, Brazil}
\author{M.~E.~S.~Pereira}
\affiliation{Department of Physics, University of Michigan, Ann Arbor, MI 48109, USA}
\author{J.~De~Vicente}
\affiliation{Centro de Investigaciones Energ\'eticas, Medioambientales y Tecnol\'ogicas (CIEMAT), Madrid, Spain}
\author{S.~Desai}
\affiliation{Department of Physics, IIT Hyderabad, Kandi, Telangana 502285, India}
\author{P.~Doel}
\affiliation{Department of Physics \& Astronomy, University College London, Gower Street, London, WC1E 6BT, UK}
\author{A.~Drlica-Wagner}
\affiliation{Department of Astronomy and Astrophysics, University of Chicago, Chicago, IL 60637, USA}
\affiliation{Fermi National Accelerator Laboratory, P. O. Box 500, Batavia, IL 60510, USA}
\affiliation{Kavli Institute for Cosmological Physics, University of Chicago, Chicago, IL 60637, USA}
\author{K.~Eckert}
\affiliation{Department of Physics and Astronomy, University of Pennsylvania, Philadelphia, PA 19104, USA}
\author{T.~F.~Eifler}
\affiliation{Department of Astronomy/Steward Observatory, University of Arizona, 933 North Cherry Avenue, Tucson, AZ 85721-0065, USA}
\affiliation{Jet Propulsion Laboratory, California Institute of Technology, 4800 Oak Grove Dr., Pasadena, CA 91109, USA}
\author{A.~E.~Evrard}
\affiliation{Department of Astronomy, University of Michigan, Ann Arbor, MI 48109, USA}
\affiliation{Department of Physics, University of Michigan, Ann Arbor, MI 48109, USA}
\author{I.~Ferrero}
\affiliation{Institute of Theoretical Astrophysics, University of Oslo. P.O. Box 1029 Blindern, NO-0315 Oslo, Norway}
\author{J.~Frieman}
\affiliation{Fermi National Accelerator Laboratory, P. O. Box 500, Batavia, IL 60510, USA}
\affiliation{Kavli Institute for Cosmological Physics, University of Chicago, Chicago, IL 60637, USA}
\author{J.~Garc\'ia-Bellido}
\affiliation{Instituto de Fisica Teorica UAM/CSIC, Universidad Autonoma de Madrid, 28049 Madrid, Spain}
\author{E.~Gaztanaga}
\affiliation{Institut d'Estudis Espacials de Catalunya (IEEC), 08034 Barcelona, Spain}
\affiliation{Institute of Space Sciences (ICE, CSIC),  Campus UAB, Carrer de Can Magrans, s/n,  08193 Barcelona, Spain}
\author{D.~W.~Gerdes}
\affiliation{Department of Astronomy, University of Michigan, Ann Arbor, MI 48109, USA}
\affiliation{Department of Physics, University of Michigan, Ann Arbor, MI 48109, USA}
\author{R.~A.~Gruendl}
\affiliation{Department of Astronomy, University of Illinois at Urbana-Champaign, 1002 W. Green Street, Urbana, IL 61801, USA}
\affiliation{National Center for Supercomputing Applications, 1205 West Clark St., Urbana, IL 61801, USA}
\author{J.~Gschwend}
\affiliation{Laborat\'orio Interinstitucional de e-Astronomia - LIneA, Rua Gal. Jos\'e Cristino 77, Rio de Janeiro, RJ - 20921-400, Brazil}
\affiliation{Observat\'orio Nacional, Rua Gal. Jos\'e Cristino 77, Rio de Janeiro, RJ - 20921-400, Brazil}
\author{G.~Gutierrez}
\affiliation{Fermi National Accelerator Laboratory, P. O. Box 500, Batavia, IL 60510, USA}
\author{W.~G.~Hartley}
\affiliation{D\'{e}partement de Physique Th\'{e}orique and Center for Astroparticle Physics, Universit\'{e} de Gen\`{e}ve, 24 quai Ernest Ansermet, CH-1211 Geneva, Switzerland}
\author{S.~R.~Hinton}
\affiliation{School of Mathematics and Physics, University of Queensland,  Brisbane, QLD 4072, Australia}
\author{D.~L.~Hollowood}
\affiliation{Santa Cruz Institute for Particle Physics, Santa Cruz, CA 95064, USA}
\author{K.~Honscheid}
\affiliation{Center for Cosmology and Astro-Particle Physics, The Ohio State University, Columbus, OH 43210, USA}
\affiliation{Department of Physics, The Ohio State University, Columbus, OH 43210, USA}
\author{D.~J.~James}
\affiliation{Center for Astrophysics $\vert$ Harvard \& Smithsonian, 60 Garden Street, Cambridge, MA 02138, USA}
\author{M.~Jarvis}
\affiliation{Department of Physics and Astronomy, University of Pennsylvania, Philadelphia, PA 19104, USA}
\author{S.~Kent}
\affiliation{Fermi National Accelerator Laboratory, P. O. Box 500, Batavia, IL 60510, USA}
\affiliation{Kavli Institute for Cosmological Physics, University of Chicago, Chicago, IL 60637, USA}
\author{K.~Kuehn}
\affiliation{Australian Astronomical Optics, Macquarie University, North Ryde, NSW 2113, Australia}
\affiliation{Lowell Observatory, 1400 Mars Hill Rd, Flagstaff, AZ 86001, USA}
\author{N.~Kuropatkin}
\affiliation{Fermi National Accelerator Laboratory, P. O. Box 500, Batavia, IL 60510, USA}
\author{M.~A.~G.~Maia}
\affiliation{Laborat\'orio Interinstitucional de e-Astronomia - LIneA, Rua Gal. Jos\'e Cristino 77, Rio de Janeiro, RJ - 20921-400, Brazil}
\affiliation{Observat\'orio Nacional, Rua Gal. Jos\'e Cristino 77, Rio de Janeiro, RJ - 20921-400, Brazil}
\author{J.~L.~Marshall}
\affiliation{George P. and Cynthia Woods Mitchell Institute for Fundamental Physics and Astronomy, and Department of Physics and Astronomy, Texas A\&M University, College Station, TX 77843,  USA}
\author{F.~Menanteau}
\affiliation{Department of Astronomy, University of Illinois at Urbana-Champaign, 1002 W. Green Street, Urbana, IL 61801, USA}
\affiliation{National Center for Supercomputing Applications, 1205 West Clark St., Urbana, IL 61801, USA}
\author{R.~Miquel}
\affiliation{Instituci\'o Catalana de Recerca i Estudis Avan\c{c}ats, E-08010 Barcelona, Spain}
\affiliation{Institut de F\'{\i}sica d'Altes Energies (IFAE), The Barcelona Institute of Science and Technology, Campus UAB, 08193 Bellaterra (Barcelona) Spain}
\author{R.~Morgan}
\affiliation{Physics Department, 2320 Chamberlin Hall, University of Wisconsin-Madison, 1150 University Avenue Madison, WI  53706-1390}
\author{J.~Myles}
\affiliation{Department of Physics, Stanford University, 382 Via Pueblo Mall, Stanford, CA 94305, USA}
\author{R.~L.~C.~Ogando}
\affiliation{Laborat\'orio Interinstitucional de e-Astronomia - LIneA, Rua Gal. Jos\'e Cristino 77, Rio de Janeiro, RJ - 20921-400, Brazil}
\affiliation{Observat\'orio Nacional, Rua Gal. Jos\'e Cristino 77, Rio de Janeiro, RJ - 20921-400, Brazil}
\author{A.~Palmese}
\affiliation{Fermi National Accelerator Laboratory, P. O. Box 500, Batavia, IL 60510, USA}
\affiliation{Kavli Institute for Cosmological Physics, University of Chicago, Chicago, IL 60637, USA}
\author{F.~Paz-Chinch\'{o}n}
\affiliation{Institute of Astronomy, University of Cambridge, Madingley Road, Cambridge CB3 0HA, UK}
\affiliation{National Center for Supercomputing Applications, 1205 West Clark St., Urbana, IL 61801, USA}
\author{A.~A.~Plazas}
\affiliation{Department of Astrophysical Sciences, Princeton University, Peyton Hall, Princeton, NJ 08544, USA}
\author{A.~Roodman}
\affiliation{Kavli Institute for Particle Astrophysics \& Cosmology, P. O. Box 2450, Stanford University, Stanford, CA 94305, USA}
\affiliation{SLAC National Accelerator Laboratory, Menlo Park, CA 94025, USA}
\author{E.~S.~Rykoff}
\affiliation{Kavli Institute for Particle Astrophysics \& Cosmology, P. O. Box 2450, Stanford University, Stanford, CA 94305, USA}
\affiliation{SLAC National Accelerator Laboratory, Menlo Park, CA 94025, USA}
\author{E.~Sanchez}
\affiliation{Centro de Investigaciones Energ\'eticas, Medioambientales y Tecnol\'ogicas (CIEMAT), Madrid, Spain}
\author{B.~Santiago}
\affiliation{Instituto de F\'\i sica, UFRGS, Caixa Postal 15051, Porto Alegre, RS - 91501-970, Brazil}
\affiliation{Laborat\'orio Interinstitucional de e-Astronomia - LIneA, Rua Gal. Jos\'e Cristino 77, Rio de Janeiro, RJ - 20921-400, Brazil}
\author{V.~Scarpine}
\affiliation{Fermi National Accelerator Laboratory, P. O. Box 500, Batavia, IL 60510, USA}
\author{M.~Schubnell}
\affiliation{Department of Physics, University of Michigan, Ann Arbor, MI 48109, USA}
\author{S.~Serrano}
\affiliation{Institut d'Estudis Espacials de Catalunya (IEEC), 08034 Barcelona, Spain}
\affiliation{Institute of Space Sciences (ICE, CSIC),  Campus UAB, Carrer de Can Magrans, s/n,  08193 Barcelona, Spain}
\author{I.~Sevilla-Noarbe}
\affiliation{Centro de Investigaciones Energ\'eticas, Medioambientales y Tecnol\'ogicas (CIEMAT), Madrid, Spain}
\author{M.~Smith}
\affiliation{School of Physics and Astronomy, University of Southampton,  Southampton, SO17 1BJ, UK}
\author{E.~Suchyta}
\affiliation{Computer Science and Mathematics Division, Oak Ridge National Laboratory, Oak Ridge, TN 37831}
\author{G.~Tarle}
\affiliation{Department of Physics, University of Michigan, Ann Arbor, MI 48109, USA}
\author{C.~To}
\affiliation{Department of Physics, Stanford University, 382 Via Pueblo Mall, Stanford, CA 94305, USA}
\affiliation{Kavli Institute for Particle Astrophysics \& Cosmology, P. O. Box 2450, Stanford University, Stanford, CA 94305, USA}
\affiliation{SLAC National Accelerator Laboratory, Menlo Park, CA 94025, USA}
\author{D.~L.~Tucker}
\affiliation{Fermi National Accelerator Laboratory, P. O. Box 500, Batavia, IL 60510, USA}
\author{T.~N.~Varga}
\affiliation{Max Planck Institute for Extraterrestrial Physics, Giessenbachstrasse, 85748 Garching, Germany}
\affiliation{Universit\"ats-Sternwarte, Fakult\"at f\"ur Physik, Ludwig-Maximilians Universit\"at M\"unchen, Scheinerstr. 1, 81679 M\"unchen, Germany}
\author{A.~R.~Walker}
\affiliation{Cerro Tololo Inter-American Observatory, NSF's National Optical-Infrared Astronomy Research Laboratory, Casilla 603, La Serena, Chile}
\author{J.~Weller}
\affiliation{Max Planck Institute for Extraterrestrial Physics, Giessenbachstrasse, 85748 Garching, Germany}
\affiliation{Universit\"ats-Sternwarte, Fakult\"at f\"ur Physik, Ludwig-Maximilians Universit\"at M\"unchen, Scheinerstr. 1, 81679 M\"unchen, Germany}
\author{W.~Wester}
\affiliation{Fermi National Accelerator Laboratory, P. O. Box 500, Batavia, IL 60510, USA}

\collaboration{100}{(The DES Collaboration)}

\begin{abstract}
  Stochastic field distortions caused by atmospheric turbulence are a
  fundamental limitation to the astrometric accuracy of ground-based
  imaging.  This distortion field is measurable at the locations of
  stars with accurate positions provided by the \gaia\ DR2 catalog; we
  develop the use of Gaussian process regression (GPR) to interpolate the distortion
  field to arbitrary locations in each exposure.  We introduce an
  extension to standard GPR techniques that exploits the knowledge that
  the 2-dimensional distortion field is curl-free.  Applied to several
  hundred 90-second exposures from the \textit{Dark Energy
    Survey} as a testbed, we find that the GPR correction reduces the
  variance of the turbulent distortions $\approx12\times$, on average,
  with better performance in denser regions of the \gaia\ catalog.  The
  RMS per-coordinate distortion in the $riz$ bands is typically
  $\approx7$~mas before any correction, and $\approx2$~mas after
  application of the GPR model. The GPR astrometric corrections are
  validated by the observation that their use reduces, from 10 to
  5~mas RMS, the residuals to an orbit fit to
  $riz$-band observations over 5 years of the $r=18.5$ trans-Neptunian object Eris.  We also
  propose a GPR method, not yet implemented, for simultaneously estimating the turbulence
  fields and the 5-dimensional stellar solutions in a stack of
  overlapping exposures, which should yield further turbulence
  reductions in future deep surveys.
\end{abstract}

\section{Introduction} \label{sec:intro}
Ground-based astrometric measurements are one of the oldest human
quantitative scientific endeavors.  The accuracy of astrometric data
were limited by the resolution of human vision, and subsequently by
the angular resolution of telescopes.  The successful detection of
stellar parallax in the mid-1800's required understanding of stellar
aberration and atmospheric refraction, which can be ameliorated to a
great extent by differential measurements with respect to nearby
more-distant stars, and attention to observing techniques
(e.g. transit telescopes).
In the modern era of
sub-arcsecond seeing and digital detectors, the error budget for
relative astrometric accuracy of unresolved sources within an exposure
will usually be 
dominated by three contributions:
\begin{enumerate}
  \item Shot noise: stochastic errors in centroiding of the source in
    pixel coordinates.  This component is typically $\sigma_x \approx
    \sigma_{\rm PSF} / \nu,$ where $\sigma_{\rm PSF}$ is the RMS width
    of the point-spread function, and $\nu$ is the signal-to-noise
    ratio (S/N) of the detection.\footnote{Throughout this paper, we
      will quote astrometric errors or image sizes as the RMS per axis
      on the sky.}
  \item Solution accuracy: the static errors in the map from pixel
      coordinates to (relative) sky coordinates, i.e. the distortions
      in the optics and detector, and the static refraction of the
      atmosphere, including chromatic distortions from the atmosphere
      and optics.
   \item Atmospheric turbulence: stochastic wander of the source due
     to refraction by atmospheric density fluctuations.
\end{enumerate}
The transfer of the relative astrometry of an image to absolute sky
coordinates is further limited by the accuracy of the reference
catalog used to make such a transfer (as well as the contributions of
items (1) and (3) to the exposure's measurements of the reference
stars).

While specialized instruments can be designed to improve solution
accuracy, we demonstrated in \citet[][B17]{decamast} that an
astrometric solution with $\approx 1$~mas accuracy is possible for a
general-purpose wide-field imager, the Dark Energy Camera (DECam)
\citep{decam} on
the 4-meter Blanco Telescope at Cerro Tololo InterAmerican
Observatory (CTIO).  This leaves shot noise (1) and atmospheric turbulence
(3) as the dominant sources of astrometric noise, with the former
dominant for faint sources and latter for high-S/N point sources.

Astrometric science has been revolutionized by space observatories,
particularly the \textit{Hipparcos} \citep{hipparcos} and \gaia\ DR2
\citep{gaiadr2} catalogs.  Space telescopes gain substantially in shot
noise errors if diffraction-limited resolution yields low $\sigma_{\rm
  PSF}$; their stability, specialized instrumentation, and greatly reduced
chromatic effects improve solution accuracy; and, perhaps most
importantly, they are free of atmospheric turbulence errors.

The advent of \gaia\ DR2 also revolutionizes the potential of
ground-based astrometry.  Most obviously, the density of the DR2
catalog ($O(1)$ star per arcmin$^2$) allows almost any ground-based
exposure of modest field of view (FOV)  to be placed onto the absolute
reference frame of \gaia\ DR2, obliterating the distinction between
absolute and relative astrometry. Typical position/parallax
uncertainties in \gaia\ DR2 rise from
$\approx0.02$~mas at $G=13$~mag to $\approx 2$~mas at the catalog
limit of $G=21$~mag \citep{gaiadr2}, 1--3 orders of magnitude lower than
previous astrometric catalogs approaching similar sky density.

In this paper we demonstrate and quantify another important benefit
that \gaia\ DR2 bestows on ground-based astrometry: the $\approx
1\arcmin$ typical spacing
between \gaia\ stars is well below the $\approx10\arcmin$ coherence
length of atmospheric turbulence, which means that we can use \gaia\ as
a reference to measure and correct the majority of the power spectrum
of astrometric distortions imposed by atmospheric turbulence.

The idea of exploiting the finite correlation length of atmospheric
turbulence to reduce the error induced on target stars'
positions has been discussed before.  This has been of particular
interest in astrometric searches for exoplanets (including use with
adaptive optics and interferometers). \citet{lazorenkos} propose a
fairly complex method to interpolate turbulence to a single target star from an
ensemble of nearby reference stars. \footnote{The method of \citet{lazorenkos} may be
  equivalent to GPR solution in some limits; we have not investigated
  this carefully.}    This method was applied to
exposures from the 8-meter VLT, yielding estimates of parallax
accuracy of 0.04~mas for stars at 17--19~mag \citep{lazorenkoFORS} for
exposures accumulating to $\approx1800$~s.  [All models and data
predict turbulence residuals to decline with exposure time as $T^{-1/2}.$]

We address in this work the application to wide-field surveys, where
we are interested in estimating positions for all targets in the
field, ideally to the photon-noise limit.  The characteristics of
atmospheric turbulence were investigated theoretically by
\citet{lindegren} and empirically by \citet{han95}
and \citet{zacharias96}, among others.
B17 found statistics for turbulent distortions to be in rough
agreement with these earlier works, and proposed the use of
Gaussian process 
regression (GPR) to transfer the turbulent field from \gaia\ stars to
targets of interest.  If the turbulence gives rise to a projected
(2-dimensional) time-delay surface that can be accurately described as
a Gaussian random field, then GPR, being the 
maximum-likelihood estimator for a Gaussian process, is also the minimum-variance unbiased
interpolator.

Earlier explorations of extraction of precision photometry from
wide-field CCD imaging in targeted fields
include \citet{PlataisMosaic}, \citet{andersonWFI}, and \citet{bouy}.
Typically, any time-dependent distortions such as atmospheric
turbulence are corrected through polynomial fits to per-exposure distortions over
the span of single CCD (scales of order 10\arcmin).
\citet{andersonWFI} take the additional step of 
referencing each star to a locally
linear transformation determined from $\approx 50$ neighboring
reference stars, obtaining $\approx7$~mas residuals in their 900~s
exposures on the 2.2-meter ESO telescope.

While this paper was under review, \citet{lubowPS1} reported
application of a similar method---local coordinate systems defined by
the 33 nearest Gaia DR2 stars---to the catalogs of the
\textit{PanSTARRS1} survey, attaining median differentials between PS1
and Gaia positions of $\approx5$~mas with exposure times of 30--45~s
on the 1.8-meter telescope.

In this paper we pursue the application of GPR astrometric
interpolation to positional catalogs from exposures in the
\textit{Dark Energy Survey (DES)} \citep{diehlDES} and develop a method that can be
applied in ``production mode'' to the $O(10^5)$ exposures and
$O(10^9)$ detections of unresolved sources in that survey.  As
reported by B17, the 90-second \des\ exposures exhibit strongly
anisotropic stochastic distortions with typical RMS amplitudes of
5--10~mas.  This dominates the $\approx1$~mas systematic errors in the 
calibration of the DECam astrometric map.  In this work we will
demonstrate that GPR from \gaia\ DR2 stars succeeds in
reducing the RMS stochastic distortions to $\approx2$~mas per axis.

This greatly surpasses the requirement set for the Vera C. Rubin Observatory of
$\le10$~mas RMS relative astrometric accuracy per axis.\footnote{See
  Table 18 of
  \url{https://docushare.lsstcorp.org/docushare/dsweb/Get/LPM-17}.} GPR
turbulence reduction will allow the \textit{Rubin Observatory Legacy Survey of Space
and Time (LSST)}\footnote{\url{https://www.lsst.org}} to push astrometric science further beyond the
capabilities of \gaia\ in many ways. The \lsst\ will be able to
measure \gaia-quality stellar parallaxes/proper motions well beyond
\gaia's faint limit, as well as improving upon \gaia\ accuracy for stars near
its limit. \lsst\ can also bring milliarcsecond precision to the tracking of
minor planets and other transients.

In the next section we review the relevant aspects of \des\
imaging and astrometric results from B17.  Section~\ref{sec:gp} 
reviews standard GPRs. Astrometric interpolation differs
from standard cases in that the turbulent image displacement $(u,v)$
is observed to follow the expectation that it is curl-free.  We show
how to extend the GPR formalism to exploit this known relation between
the $u$ and $v$ fields.

Section~\ref{sec:kernel} derives the correlation function---or, in GPR
parlance, the ``kernel''---that should result from wind-blown
von Karman turbulence at a single layer of the atmosphere.
Sections~\ref{sec:methods} describes the
numerical methods for choosing and applying a kernel to the catalogs,
and Section~\ref{sec:results} gives quantitative results of
application of the curl-free GPR to a test sample of several hundred
\des\ exposures, including validation by fitting an orbit to \des\
observations of the bright trans-Neptunian object Eris.   We conclude
in Section~\ref{sec:conclusion}.

In Appendix~\ref{sec:bigsolution}, we derive an even more comprehensive use
of the GPR methodology, in which one can simultaneously obtain the
maximum-likelihood values for the distortion fields of a stack of
exposures \emph{and} the 5d position/parallax/proper-motion solutions
of the stars contained within this stack.  This method has the
potential for significant further reduction of turbulence residuals,
by effectively turning every high-S/N star in the field into a
reference star, not just those with \gaia\ measures.  We have not yet
implemented this method on \des\ data.

\section{Summary of DES data and astrometry}
\subsection{\des\ data and astrometric solution}
The DECam science array consists of 62 distinct CCDs, each
$2048\times4096$ pixels at $\approx0\farcs263$ per pixel.  The field
of view (FOV) approximates a circle with $\approx1\arcdeg$ radius.  
The analyses in this paper are done on exposures taken as part of the
``Wide'' survey of \des, in which a 5000~deg$^2$ section of the
southern galactic cap is imaged 10 times in each of the $g,r,i,z,$ and
$Y$ filter bands, spread over 6 annual Aug-Feb observing seasons.
We will for the most part ignore the $Y$-band exposures, which have
substantially lower $S/N$ than $griz$ and will not contribute
substantially to overall astrometric precision (but appear otherwise
astrometrically well-behaved).  The $griz$ exposures for the Wide
survey are all 90~s duration.

The results reported here make use of the ``Y6A1'' internal release of
the full Wide survey data.  The individual exposures are processed
with the ``FinalCut'' pipeline very similar to the earlier version
described in \citet{ y1desdm}.  Pixel coordinates for all sources are
determined from the \texttt{[XY]WIN\_IMAGE} windowed centroid quantity
 measured by \texttt{SExtractor} \citep{sextractor}.  We assign the
\texttt{ERRAWIN\_IMAGE} measurement as the $\sigma$ of a circular
Gaussian measurement error on each unresolved source.

A mapping from pixel coordinates to sky coordinates is derived using
the methods described in B17, with some improvements.  The astrometry
solution includes these terms:
\begin{itemize}
  \item A cubic polynomial spanning the whole FOV for each exposure,
    which absorbs the 
    static atmospheric refraction, stellar aberration, and a pointing solution.
  \item A zenith-oriented differential chromatic refraction term.
  \item A polynomial per CCD per band per observing season, which
    captures optical distortions.
  \item A chromatic lateral color shift oriented radially.
  \item $\ast$Short-time-scale (weeks) affine shifts in the
    positions of the CCDs in the focal plane.
  \item The ``tree ring'' and ``glowing edge'' distortions arising
    from stray electric fields in each CCD.
  \item $\ast$An additional fixed map of distortions apparently due to
    electric fields around the electrical cable connector.
  \end{itemize}
  Those items marked with ($\ast$) have been added to the model since
  B17.  Furthermore, the entire astrometric solution is now registered
  to \gaia\ DR2 and allows for nonzero proper motion and parallax for
  all stars in the DES footprint when registering images internally
  and to \gaia.  In B17 it was demonstrated that any errors in the
  astrometric solution that repeat over time are limited to
  $\lesssim1$~mas RMS.

\subsection{Nature of the stochastic distortions}
Aside from the per-exposure cubic polynomial spanning the full FOV,
these baseline Y6A1 astrometric solutions do not attempt to remove any
of the stochastic distortions that would arise from atmospheric
turbulence or other effects varying on timescales of single
exposures.  B17 described several properties of the stochastic distortion
patterns, which we summarize here.

We quantify the stochastic distortions primarily through the 2-point
correlation functions of the residuals $u$ and $v$ between the
\des-derived RA/Dec and the \gaia~DR2 values.  We define
\begin{equation}
  \xi_{uu}(\vx) = \langle u_i u_j \rangle,
\end{equation}
where $i$ and $j$ range over all stars separated by the vector \vx.
The virtue of this statistic is that the contribution from shot
noise---or any other form of noise that has negligible star-to-star
correlation---averages to zero.  We can similarly define $\xi_{vv},
\xi_{uv}.$ Of particular interest is
\begin{equation}
\label{xiplus}
\xi_+(\vx) \equiv \xi_{uu}(\vx) + \xi_{vv}(\vx).
\end{equation}
As $\vx\rightarrow0$, the value of $\xi_{uu}$ yields the total RMS
variance in the $u$ direction caused by atmospheric turbulence (or
other spatially correlated errors).

\begin{figure}[tb]
	\centering
	\includegraphics[width=\columnwidth]{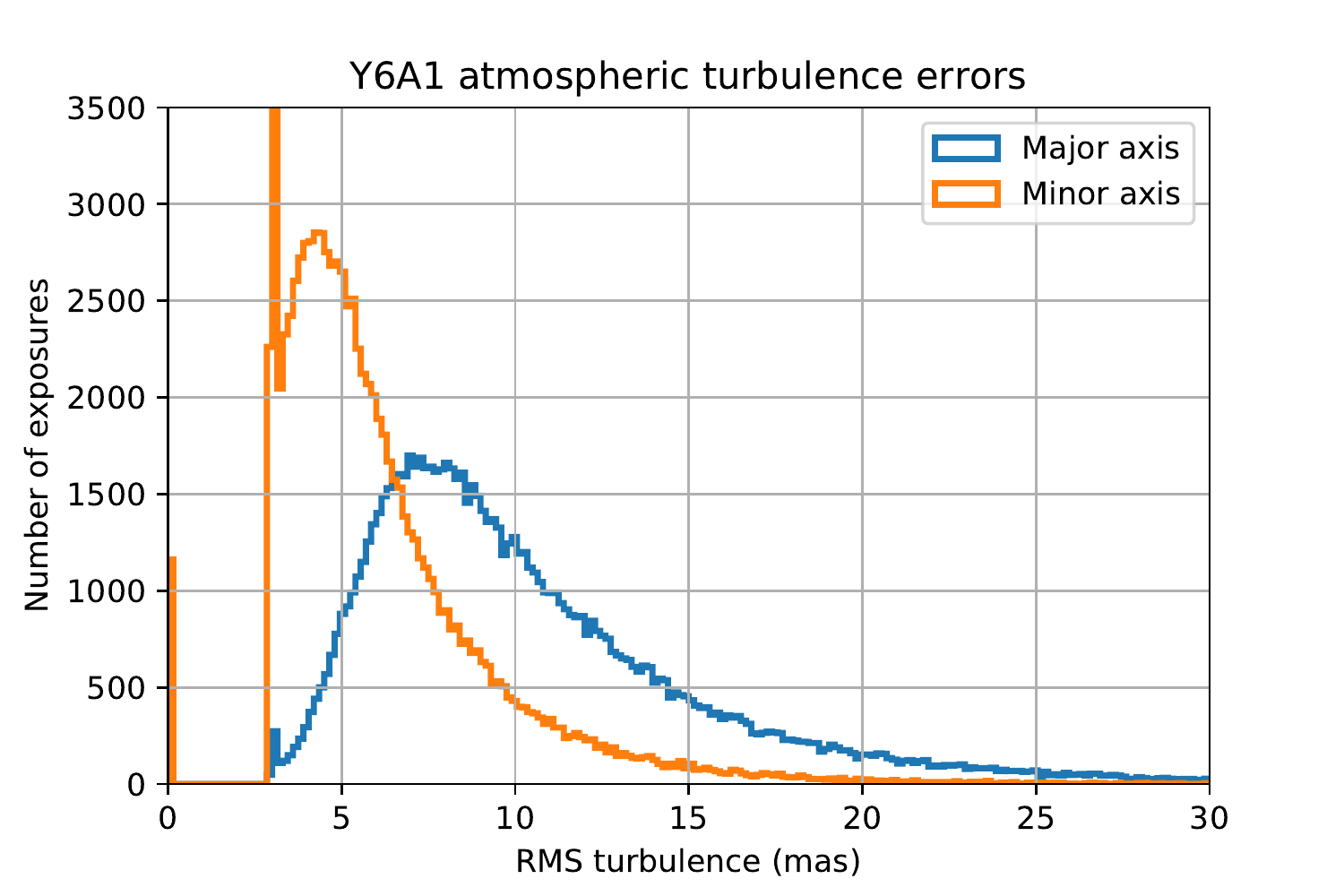}
	\caption{Distribution of the major and minor axes of the
          turbulence-induced astrometric errors for the
          $\approx80,000$ exposures of the \des\ Y6A1 Wide survey.
          The distributions are clipped at 3~mas since
          smaller values are unreliable in these data.}
\label{majorminor}
\end{figure}
 
The \des\ exposures are seen to have anisotropic errors, \ie\
$\xi_{uu}\ne\xi_{vv}$ and $\xi_{uv}\ne0.$  Figure~\ref{majorminor}
shows the distribution of major/minor axes of the error ellipse
implied by the zero-lag limits of the $\xi$'s.  The modal major and minor
axes are 7~mas and 5~mas, respectively, with the means being higher.
This is the turbulence noise.

\begin{figure}[ht]
\centering
\includegraphics[width=\columnwidth]{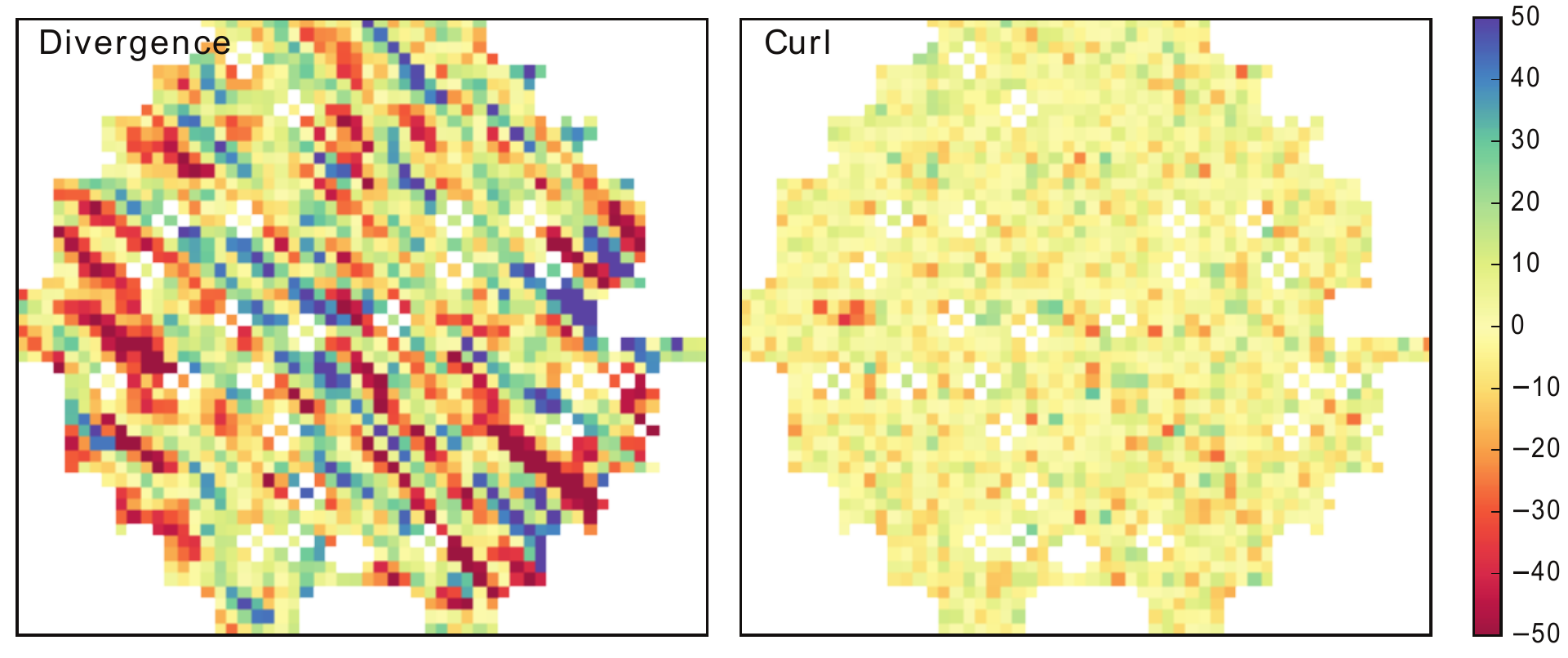}
\caption{The divergence and curl of the astrometric residuals 
  on exposure \#228645 (30~s, $z$ band) are plotted on a
  common scale.  
The continuity of the vector field across chip boundaries, the curl-free nature of the
  field, and the streaky pattern of divergence suggests the
  hypothesis that these distortions arise 
  from wind-blown atmospheric turbulence. Reproduced from BK17.}
\label{ebmap}
\end{figure}
 A further critical observation is that the stochastic distortion field
is curl-free, as seen in Figure~\ref{ebmap}.  This implies that the
scalar $\xi_+(\vx)$ function is a complete description of the vector turbulent
distortions, if they arise from a Gaussian process, as explained in
the Appendix of BK17.  As will be
detailed below, this function indicates a typical correlation length
of $\approx6\arcmin$ for
the stochastic distortions (where $\xi_+$ drops to half its $\vx=0$ value)

\section{Curl-free, anisotropic Gaussian process interpolation}
\label{sec:gp}
\subsection{Scalar Gaussian process review}
The standard methodology for Gaussian process regression (GPR) follows
from the assumption that we are interested in the value of some
stochastic scalar $u(\vx)$ over the field $\vx$ (in our case, the
positions $\vx=(x,y)$ of a star on a local projection of the sky). The
generation of $u$ is considered a Gaussian process if and only if the
distribution of $u$ at any collection of points $\vx = \{\vx_1,
\vx_2,\ldots,\vx_N\}$ can be described by a multivariate normal
distribution\footnote{Using the standard notation $\mathcal{N}(\mu,C)$
  for a multivariate normal distribution with mean $\mu$ and
  covariance matrix $C$.}
\begin{equation}
  \vu \equiv \{u_1,u_2,\ldots,u_N\}  \sim \mathcal{N}(\boldsymbol{\mu},
  \matK)
\end{equation}  
with a mean $\boldsymbol{\mu}$ that we can usually take as zero, and a
 covariance matrix $\matK\equiv\langle\vu \vu^T\rangle$ with a known form $K_{ij}=K(\vx_i,\vx_j)$  We
 will consider stationary (but potentially anisotropic) fields with
 possible hyperparameters $\vpi_K,$ for
 which we can place $K_{ij}= K(\vx_i-\vx_j; \vpi_K).$

 If we consider the field to have known values $\vu$ at training
 points $\vx$, and we seek estimates $\vu^\star$ at points
 $\vx^\star$, then we can write the joint distribution as
 \begin{equation}
   \left(\begin{array}{c}
           \vu \\
           \vu^\star
         \end{array}\right)
   \sim \mathcal{N}\left( 0,
     \left[ \begin{array}{cc}
               \matK & \matK^\star \\
               \matK^{\star T} &  \matK^{\star\star}
             \end{array}
      \right]
  \right).
 \end{equation}
Here we follow GP convention by setting $K^\star_{ij} =
K(\vx_i-\vx^\star_j)$ and $K^{\star\star}_{ij} =
K(\vx^\star_i-\vx^\star_j).$
Conditioning this joint distribution on the measured \vu\ yields the
standard scalar GPR solution:
\begin{equation}
  \vu^\star | \vu \sim \mathcal{N}\left( \matK^{\star T} \matK^{-1} \vu,
    \matK^{\star\star}-\matK^{\star T} \matK^{-1} \matK^\star \right).
  \label{gpr1}
\end{equation}
Thus the maximum-likelihood (ML) solution is the mean of this
  conditioned distribution---the ML value of $\vu^\star$ at any point is a
  linear combination of the values at the \vu, with
  coefficients determined by the relative locations of the reference
  stars to the target position.
The presence of independent measurement noise $\vn=\{n_1,n_2,\ldots\}$ on the
training points can be incorporated by considering the noise to be an
addition to the kernel / covariance matrix \matK\ that has zero
correlation length, \ie\ only appears on the diagonal
\begin{equation}
  K_{ij} \rightarrow K(\vx_i-\vx_j) + \delta_{ij} \left\langle n_i^2
\right\rangle.
\end{equation}
The suitability of the chosen kernel to the training data is
quantified by the log marginal likelihood,
\begin{equation}
  -2\log p(\vu | \vx, \vpi_K) = \vu^T \matK^{-1} \vu + \log |2\pi\matK|,
\label{lml}
\end{equation}
where $\matK$ is a function of the hyperparameters $\vpi_K$.  The most
suitable kernel is chosen by maximizing this quantity over $\vpi_K$.
If the field is truly Gaussian, the resultant interpolator is then
optimal.  The GPR does, however, yield a functional interpolator even if
\matK\ is not precisely the covariance function of $u$, or if
$u$ is not a Gaussian field.

Application of the standard GPR to ground-based astrometric data
proceeds by using the sky positions of the stars as the feature
vectors $\vx_i$, and defining two distinct scalar fields $(u,v)$ as
the differences between the observed $x$ and $y$ coordinates and the
true coordinates, $(u_i,v_i) = \vx^{\rm obs}_i - \vx^{\rm true}_i.$
The training data are those for which the \gaia~DR2 motions are
available (and calculated at the observation epoch).   Distinct GPR's
are trained for each component $u$ and $v$ in this simplest scheme.

\subsection{Curl-free vector fields}
\label{curlfree}
A simple and applicable ray-optic model for the displacements caused
by atmospheric turbulence is that the centroid of the stellar image moves
by an apparent vector
\begin{equation}
  \vu = (u,v)= \boldsymbol{\nabla}_x \phi(\vx)
\label{gradient}
\end{equation}
where $\phi(\vx)$ is the optical path difference (OPD) along the line of
sight $\vx$, as convolved with the telescope aperture.  If this model
holds instantaneously, and the source photon arrival rate is constant,
then we can average both $\vu$ and $\phi$ over the duration of the
exposure and the equation will still hold. A special case would be
a single-screen ``frozen''
approximation: $\phi(\vx, t)$ varies with time as
$\phi(\vx - \boldsymbol{w}t)$ for a wind vector $\boldsymbol{w}$ at the
turbulent layer.  In this case, the time-averaged $\phi$ is the
convolution of the instantaneous $\phi$ with a line segment of wind
motion during the exposure.  The results in this section will,
however, be true for any model of the time-averaged $\phi.$

As shown in B17, the two-point correlation functions
of the
residual displacement fields $\vu(\vx)$ are observed to be purely ``E-mode,'' meaning
that $\boldsymbol{\nabla\times}\vu$ contains only white noise
consistent with the shot noise of stellar centroid measurements.
It is therefore likely that enforcing
a curl-free turbulence field will yield a more accurate GPR, since we
can combine the data from both fields $u$ and $v$ to solve for a
single degree of freedom $\phi.$

The curl-free GPR begins by assuming that $\phi$ is a zero-mean
Gaussian field, in which case it can be fully characterized by its
power spectrum $P_\phi(\vk)$ or its two-point correlation function
$K_\phi(\Delta\vx)$, which we can relate via the Fourier transform
\begin{equation}
  K_\phi(\Delta\vx) = \int d^2\vk\, P_\phi(\vk) e^{i\vk\cdot\Delta\vx}.
\end{equation}  
The relation (\ref{gradient}) implies that the Fourier transforms of
$\vu$ and $\phi$ satisfy $FT(\vu) = i\vk\, FT(\phi).$  This in turn
implies that the covariance matrix of the turbulence
component of the displacement field is
\begin{equation}
  K_t(\Delta \vx) = \int d^2\vk\, \left(\vk\vk^T\right)P_\phi(\vk)
  e^{i\vk\cdot\Delta\vx}.
\label{Kft1}
\end{equation}
Thus any parametric model for $P_\phi$ can be transformed into a
parametric $2\times2$ covariance matrix function of $\Delta\vx.$  
[We retain a scalar notation for this $2\times2$ function $K_t,$
reserving the matrix symbol $\matK$ for the
matrix of covariances between different field points.]

With this modification, we can recast the GPR to
a merged vector $\vU = \{u_1,v_1, u_2, v_2, \ldots,
u_{N_s},v_{N_s}\}$ of $2N_s$ scalar quantities.  As in the scalar
GPR, we can define matrices $\matK, \matK^\star, \matK^{\star\star}$
composed of $2\times2$ blocks such that
\begin{align}
  K_{i,j} & = K_t(\vx_i-\vx_j) + \delta_{ij} \left\langle \vn_i \vn_i^T
  \right\rangle \nonumber \\ 
  K^\star_{i,j} & = K_t(\vx_i-\vx^\star_j)  \label{k2d} \\
  K^{\star\star}_{i,j} & = K_t(\vx^\star_i-\vx^\star_j). \nonumber
\end{align}
The measurement noise term in the first row is also a $2\times2$
matrix, which in practice is very close to diagonal for \des\ data but
is substantially anisotropic for \gaia.
After this change, the standard GPR formulae (\ref{gpr1}) and
(\ref{lml}) remain correct, with the substitution $\vu\rightarrow\vU.$
If the $K_t(\Delta\vx)$ function were
diagonal, then the GPR solution would separate into two distinct
scalar GPRs.  But even an isotropic field $\phi$ has off-diagonal
terms in $K_t$ and therefore this curl-free interpolant exploits
information that two distinct scalar GPRs cannot.

\section{Model for wind-blown von Karman turbulence}
\label{sec:kernel}
We desire a model for the correlation function $K(\vx)$ between the
turbulent distortions at two locations separated by \vx.  Since the
displacement $(u,v)$ is 2-dimensional, there are three scalar
correlation functions of interest, $K_{uu}, K_{uv},K_{vv}.$
Considering first one dimension of the deflection, $u$, we have from
\eqq{gradient} that the deflection of a single photon passing through
a time-delay screen $\phi$ at location \vx\ and time $t$ will be
$u \propto \frac{\partial}{\partial x} \phi(\vx,t).$   The photons that
arrive to form the image of a particular star arrive at fixed $t$ from a range of \vx\ described by the aperture function of the
telescope $A(\vx)$, effectively convolving the instantaneous OPD
screen as $\phi\rightarrow \phi \otimes A.$ Adopting the frozen-screen
approximation common in adaptive-optics analyses, such that 
$\phi(\vx,t)=\phi(\vx - \boldsymbol{w}t)$, means that the integration
of photon arrival time over the exposure duration $T$
corresponds to a further
convolution of the phase screen by a line segment of length and
direction $\boldsymbol{w}T,$  which we will denote by the ``wind
function'' $W(\vx).$ The apparent deflection, averaged over all
arriving photons for a star crossing the phase screen at location
$\vx,$ becomes
\begin{equation}
  u(\vx) \propto \frac{\partial}{\partial x} \left[ \phi(t=0)
    \otimes A  \otimes W\right].
\end{equation}
The convolutions become multiplications in the Fourier domain, and we
can also recall that the correlation function $K_{uu}$ is the Fourier
transform of the power spectrum $P_u(\vk)$ of the deflections, giving
\begin{multline}
  \left( \begin{array}{c}
           K_{uu}(\vx) \\
           K_{uv}(\vx) \\
           K_{vv}(\vx)
         \end{array} \right) \propto \\
       \int d^2k\,  \left( \begin{array}{c}
           k_x^2 \\
           k_xk_y \\
           k_y^2
         \end{array} \right) P_\phi(\vk) \left|\tilde A^2(\vk)\right|
       \left|\tilde W^2(\vk)\right| e^{i\vk \cdot \vx},
       \label{Kft2}
     \end{multline}
     which reproduces \eqq{Kft1} by the inclusion of the aperture and wind
functions altering the instantaneous OPD screen $\phi.$  In this
section we have treated \vx\ as being in distance units and \vk\ in
inverse distance.  But we can equally well use these equations with
\vx\ representing the 2d angle subtended about the telescope axis,
which we shall do henceforth.

It now remains to choose models for $P_\phi, A,$ and $W$.  The common
model for the power spectrum of atmospheric turbulence is the von
Karman spectrum,
\begin{equation}
  P_\phi(k) \propto \left(k^2 + k_0^2\right)^{-11/6},
  \label{vonkarman}
\end{equation}
where $k_0$ is the inverse of the outer scale of turbulence.  More
precisely, our angular system, with turbulence at height $h$ and
outer scale $r_0$, produces $k_0=2\pi h / r_0.$  There is an additional
complication that, when the zenith angle $z$ of the observation is
nonzero, the (horizontal) turbulent
layer is foreshortened along the direction toward zenith.  To include
this effect, we should break \vk\ into components $(k_\parallel,
k_\perp)$ relative to the parallactic direction, and substitute
\begin{equation}
  k^2 \rightarrow (k_\parallel\cos z )^2 + k_\perp^2
  \label{foreshorten}
\end{equation}
in the argument of $P_\phi$.  

For the telescope aperture we will adopt the simplest case of a
uniformly filled circular aperture, which yields
\begin{equation}
  \left|\tilde A^2(\vk)\right| \propto \left[
    \frac{J_1(kd/2)}{kd}\right]^2,
\end{equation}
where $J$ is the Bessel function of the first kind, and where $d$ is the
angular size of the telescope diameter $D$ as projected to the
turbulent layer, $d=D/h.$

The Fourier transform of the line-segment window function yields
\begin{equation}
\left|\tilde W^2(\vk)\right| \propto \textrm{sinc}^2\left(\vk\cdot\boldsymbol{w}/2\right),
\end{equation}
where $\boldsymbol{w}$ is now taken to be the wind transport over the
duration of the exposure, as projected onto the angular coordinates of
the telescope.

Combining the previous 5 equations yields our model for the
astrometric correlation function $\matK(\vx)$ for a single layer of
von Karman turbulence.  There are two known parameters (the zenith angle
$z$ and the parallactic angle defining $k_\parallel$) and five free
parameters: $\{\xi_0, r_0, d, w_x, w_y\}$ denoting the overall
amplitude of the turbulence, $\xi_0=(K_{uu}+K_{vv})(\vx=0)$, the
(angular equivalents of) outer scale, telescope aperture, and
components of the wind vector.  Of these parameters the outer scale
$r_0$ is the least important because it is typically larger than the
4-meter telescope diameter, and the aperture function damps the power
spectrum before the outer scale sets in.

\begin{figure}[ht]
  \centering
  \includegraphics[width=\columnwidth]{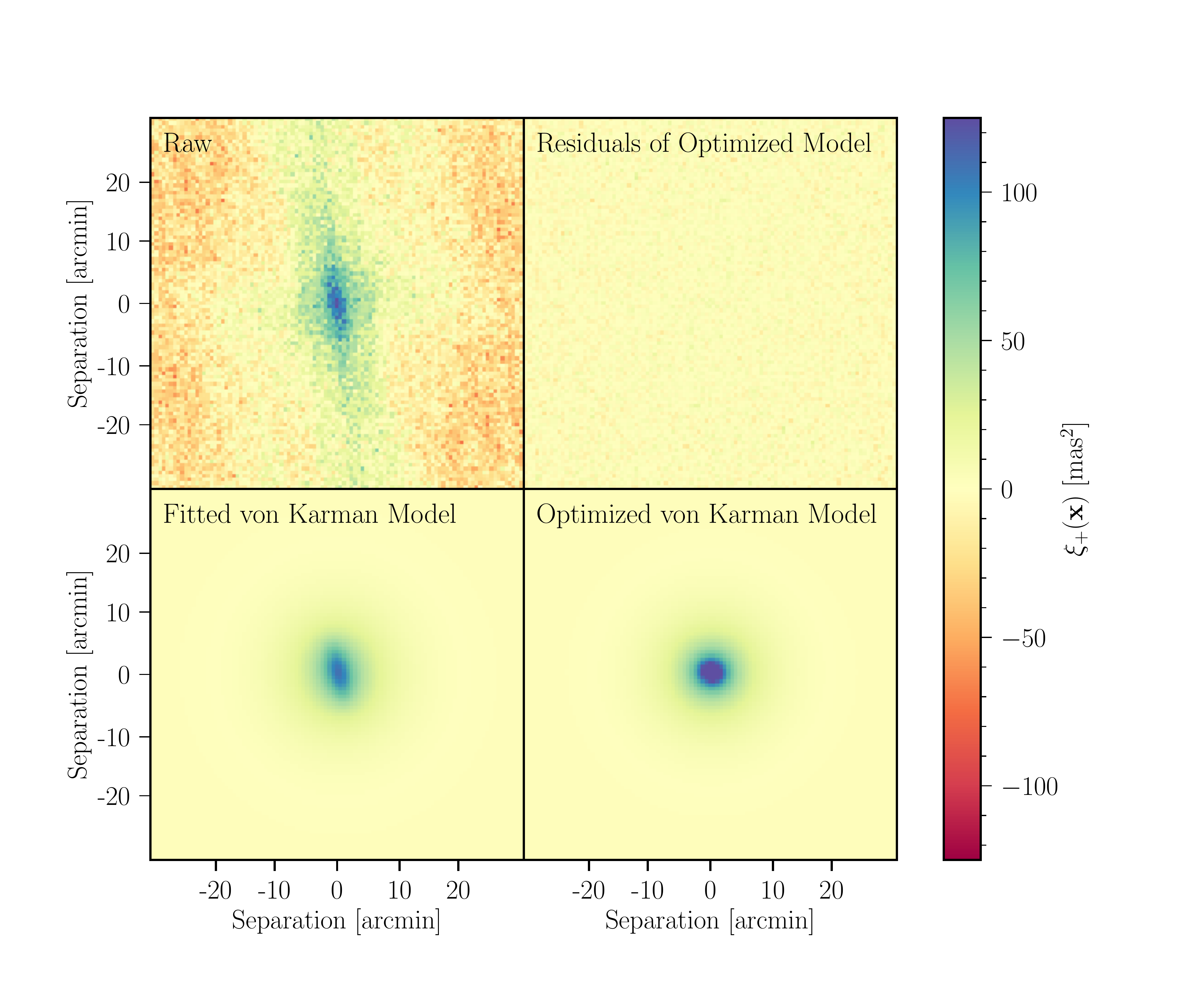}
  \caption{Each panel plots a measure or model of the astrometric error correlation function
$\xi_+(\vx).$ The upper left shows the measured correlation function
for exposure 361582 (in $i$ band), before any GPR subtraction.  At
lower left is the single-screen wind-blown von Karman turbulence model
that is the best fit to these data for $|\vx|<5\arcmin.$  At lower
right is the von Karman model that the optimizer finds to minimize the
$\xi_0$ after GPR subtraction.  At upper right is the measured $\xi_+$
after subtraction of the GPR.  Note that there is no noticeable
remaining correlation at any \vx\ (at this dynamic range) after the GPR
correction is applied.}
  \label{fig:xi2d}
\end{figure}

The real-space 2d correlation functions for the von Karman spectrum,
the aperture, and the wind function can be done analytically
individually, but their convolution is not analytic.  One has the
option of doing the convolution numerically, but we opt instead to
use the above equations to multiply the three (analytic)
Fourier-domain functions, and do the Fourier transform numerically
using an FFT.  The lower panels of Figure~\ref{fig:xi2d} plot $\xi_+=K_{uu}+K_{vv}$
for two variants of the wind-blown von Karman model.

We note at this point that the real atmospheric turbulence is unlikely to
  arise from a single von Karman screen.  The observed $\xi_+$
  functions have more structure than the von Karman model, 
  e.g. features at multiple position angles.  The turbulence may arise
  at multiple heights with varying wind speed, for example.  We will
  however proceed with this model as a viable kernel, and remember
  that it is possible that other kernels would perform better, and the
  most successful von Karman parameters may not represent the physical
  conditions of a single screen.  Data from instruments monitoring
  atmospheric turbulence is available at CTIO, but we do not expect
  \textit{ab initio} information on the turbulence parameters to
  provide a more effective GP kernel 
  than fitting and optimization to the empirical information from Gaia
  stars' residuals.

\section{Numerical methods}
\label{sec:methods}
The curl-free GPR is implemented on the \des\ data on an exposure-by-exposure basis, with the following procedure.
\begin{enumerate}
    \item \label{item:dataretrieval} \textbf{Data Retrieval}: A Python
      program retrieves the Y6A1 \des\ single-epoch astrometric
      solutions for a single exposure from \des\ data files, including estimates of shot noise $n_i$ from the \texttt{SExtractor} quantity (\texttt{ERRAWIN\_WORLD}). Corresponding \gaia\ 5-parameter astrometry solutions, and their full covariance matrices, are retrieved via the Python package \textsc{AstroQuery}.
    
    \item \label{item:preprocessing} \textbf{Pre-processing}: The retrieved \gaia\ 5d solution for epoch \texttt{J2015.5} for each star is reduced to a 2d position (and uncertainty) at the time of the \des\ exposure. The \des\ and \gaia\ catalogs are then matched with the \texttt{match\_coordinates\_sky} routine from \textsc{Astropy}; detections within 0.5\arcsec are matched. \gaia\ detections with no \des\ match are discarded. Additionally, if more than 15000 matches are found, 15000 detections are randomly chosen and the rest are discarded in order to conserve memory in later steps. The program then performs a gnomonic projection on both catalogs using the center of the \des\ exposure as the projection axis.  For each matched star, the residual field $\vu=\vx_{\textit{DES}}-\vx_{\textit{Gaia}}$ is formed, and a $2\times2$ shot-noise uncertainty  $\textbf{n}_i$ is created from the sum of the \des\ and \gaia\ measurement errors.
    
    In order to facilitate 5-fold cross-validation later on, the matched data is randomly distributed into five subsets, A through E. At any one time, one subset will be the validation set, and the other four will be the training set. Additionally, the detections in the \des\ catalog that did not have a match are designated as the prediction set.

    \item \label{item:errorcorrecting} \textbf{Error Rescaling}: In order to remove outliers, sigma-clipping is performed to four standard deviations on the residual values, $\vu_i$. Additionally, any detections that have shot-noise errors $>250$ mas$^2$ are removed. For $r$, $i$, and $z$-band exposures there are few detections with shot-noise errors this high, but this is an important step for the $g$ and $Y$ bands. We also find that the shot-noise errors from \texttt{SExtractor} are systematically underestimated, particularly for detections with very low shot noise.  To give the kernel more accurate shot-noise values, we replace the  (\texttt{ERRAWIN\_WORLD}) with the RMS 
of the residual field values for a group of 256 detections with similar estimated shot-noise.
    
The best-fit third-degree polynomial as a function of astrometric
position in the FOV is subtracted from the residual field. This is done to remove large-scale systematic errors from instrument distortion (e.g., thermal expansion of the telescope) as well as low-altitude ``ground-layer" turbulence that distort large angular scales.

    \item \label{item:fitting} \textbf{Fitting kernel to $\xi_+$}: In order to
      get an initial guess of the kernel parameters, $\vpi_K$,
      $\xi_+(\vx)$ is calculated for the training set. A Nelder-Mead
      optimizer varies $\vpi_K$ to minimize the residual sum of
      squares (RSS) between the observed $\xi_+$  and the parametric
      von-Karman $\xi_+=(K_{uu} + K_{vv})$. For this least-squares
      fitting,  only $\xi_+(\vx)$ for $x<5\arcmin$ is used.
    
      With the resultant kernel parameters, we execute the curl-free GPR on the $\vu_i.$ Using 5-fold cross-validation, we obtain a residual $\tilde\vu_i$ for each matched star from the difference of the raw $\vu_i$ and a GPR-estimated value that has \textit{not} been trained on that star.  Another round of sigma-clipping is then performed on the data, this time removing all detections greater than four standard deviations from the mean of $\tilde{\vu}_*$.

    \item \label{item:optimization} \textbf{Optimization of kernel}:
      Ideally the GPR kernel fitted to the observed $\xi_+$ function
      would be the optimal interpolator, yielding the smallest RMS
      errors in position for the validation set.  The RMS errors
      can be measured without shot-noise biases as the limit of
      $\xi_+(\vx)$ as $\vx\rightarrow 0.$ We define a figure of merit
    meant to approximate this limit as
      \begin{equation}
        \xi_{1.2} \equiv \langle \xi_+(\vx) \rangle_{|\vx|<1\farcm2},
      \end{equation}
      where the average is pair-weighted.

      In practice the fitted kernel is not optimal for $\xi_{1.2},$
      perhaps because the von Karman kernel model does not
      fully describe the the field, or perhaps because of
      non-Gaussianity and/or outliers in the data.  Regardless of the
      cause, we find that the figure of merit $\xi_{1.2}$
      can be reduced
      by further optimization of the von Karman parameters $\vpi_K.$
Using the ``fitted'' kernel parameters found in the previous step as
an initial guess, we use the \texttt{L\_BFGS\_B} gradient-descent
method of the \texttt{scipy.optimize.minimize} 
optimizer \citep{SciPy} to minimize the $\xi_{1.2}$ of the validation set. 

This is the most computationally intensive part of the process since
the optimization requires many repeated evaluations of the GPR.
In order to reduce the number of optimization steps, we first optimize
over the parameters $d$, $w_x$, and $w_y,$ then fix these and optimize
the remaining parameters $\xi_0$ and $r_0$.

The results $\vpi_K$ we call the ``optimized'' kernel parameters.
We evaluate the GPR and residuals using the optimized
parameters. Again using 5-fold cross-validation and four standard
deviation sigma-clipping, we find $\tilde{\vu}_*$ for all matched
stars.
When we refer to ``GPR-subtracted'' data, it will by default
  mean those corrected with the optimized parameters.  In some of the
  further analyses we will explicitly compare results with a GPR using
  the fitted parameters from step (4) vs the optimized parameters from
  this step.

\item \textbf{Analysis}: With $\tilde{\vu}_*$, we calculate statistics on how well the model performs.  All statistics omit the stars which have been sigma-clipped.
\end{enumerate} 

\section{Results for DES exposures}
\label{sec:results}
We applied this procedure to 343 \des\ exposures (76, 70, 64, 72, and 61 from $grizY$ bands, respectively). The $g$-band exposures exhibit very high outlier rates and  \texttt{ERRAWIN\_WORLD}  inaccuracies for reasons that are not understood and probably irrelevant to the turbulence estimation, so we will not make further use of them in this paper.  The low $S/N$ of the $Y$-band exposures makes them less interesting tests of turbulence reduction, so most of our statistics will concentrate on $riz.$

The average density of matches with \gaia\ (after sigma-clipping) for the exposures analyzed is 0.95~arcmin$^{-2}$, or $10^4$ per 3~deg$^2$ \des\ exposure.  With the 5-fold cross-validation, there are an average of $\approx2000$ stars in a validation subset.
The optimization of  the kernel parameters using the
\texttt{L\_BFGS\_B} algorithm requires $\approx100$ re-calculations of
the GPR on average, which is the computational bottleneck of the procedure, requiring several hours on a 12-core cluster node.  Further investigation of the optimization may yield substantial speed-ups.

\begin{figure*}[ht!]
  \centering
  \includegraphics[width=\textwidth]{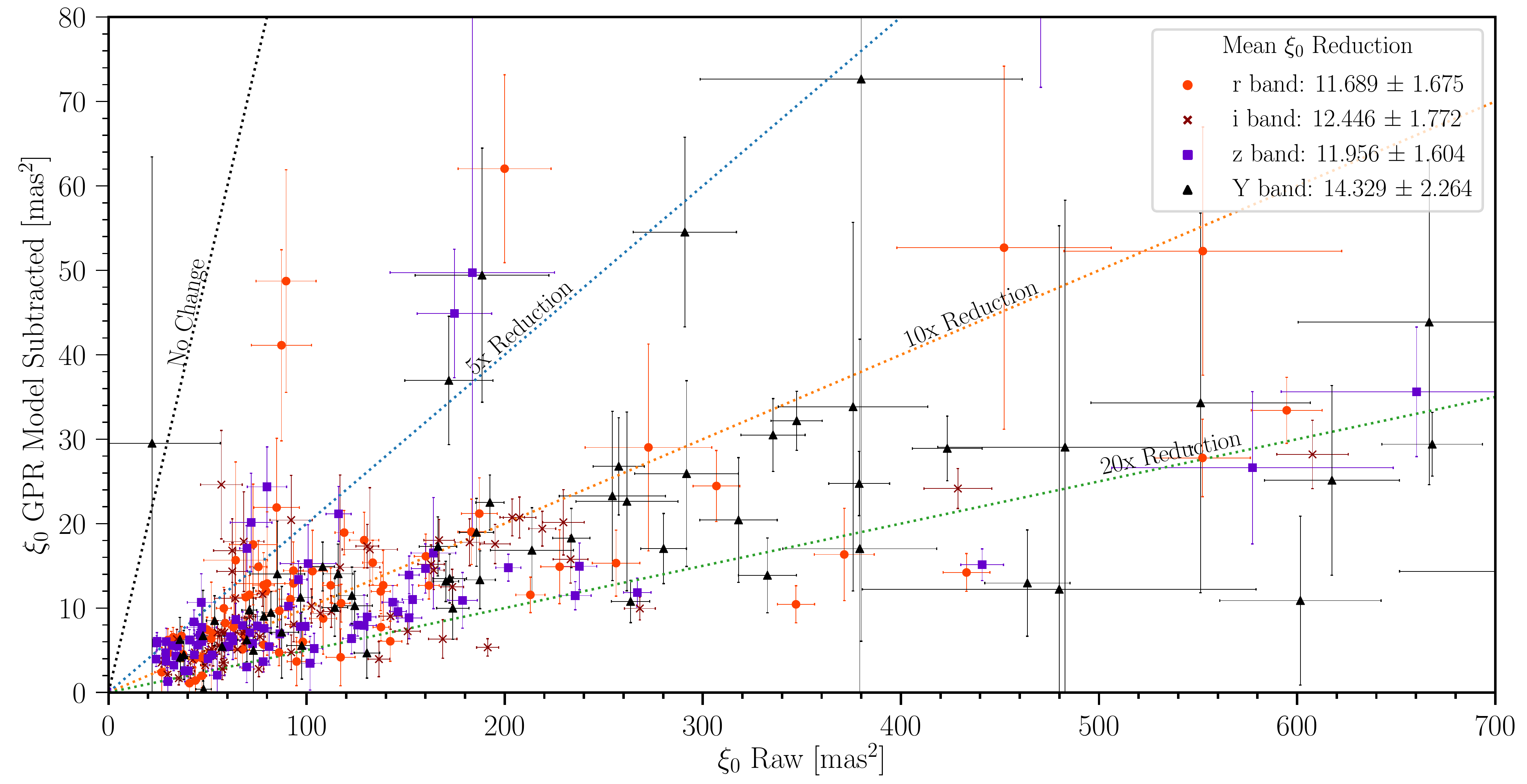}
  \includegraphics[width=\textwidth]{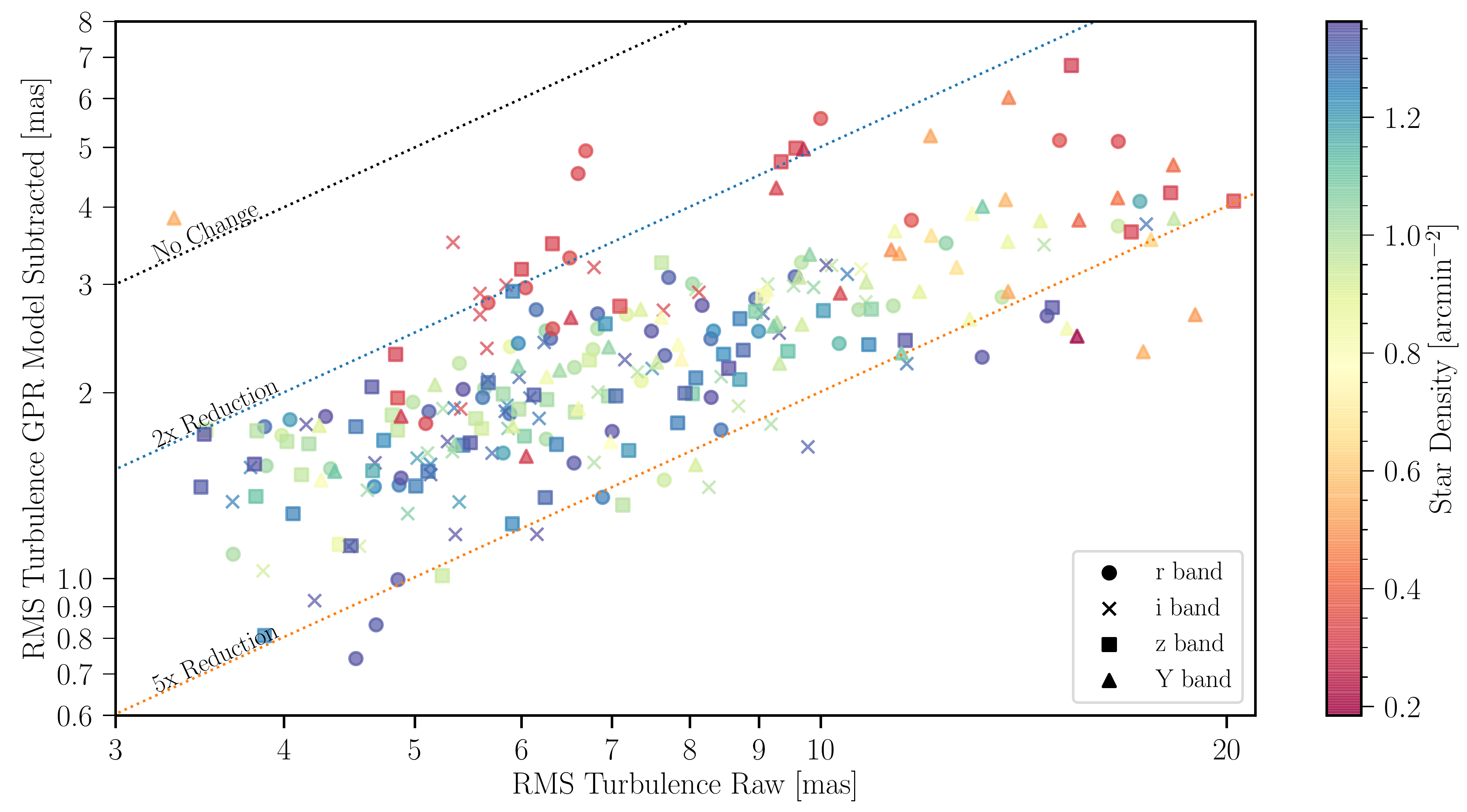}
  \caption{The upper panel plots the values of $\xi_0 \approx
    \xi_+(\lvert \vx \rvert < 0.5\arcmin)$ for the validation samples
    of every analyzed $rizY$  exposure before ($x$ axis) vs after ($y$
    axis) subtraction of the GPR.   The dotted lines mark constant
    factors of $\xi_0$ reduction by the GPR subtraction.  The lower
    plot is similar except that it uses $\sqrt{\xi_0/2},$ the RMS
    turbulence error per axis, instead of the total variance $\xi_0,$
    on the axes.  And the color scale in this case indicates the
    density of \gaia\ stars used in each exposure.  }
  \label{fig:xiBA}
\end{figure*}

\begin{figure}[ht]
  \centering
  \includegraphics[width=\columnwidth]{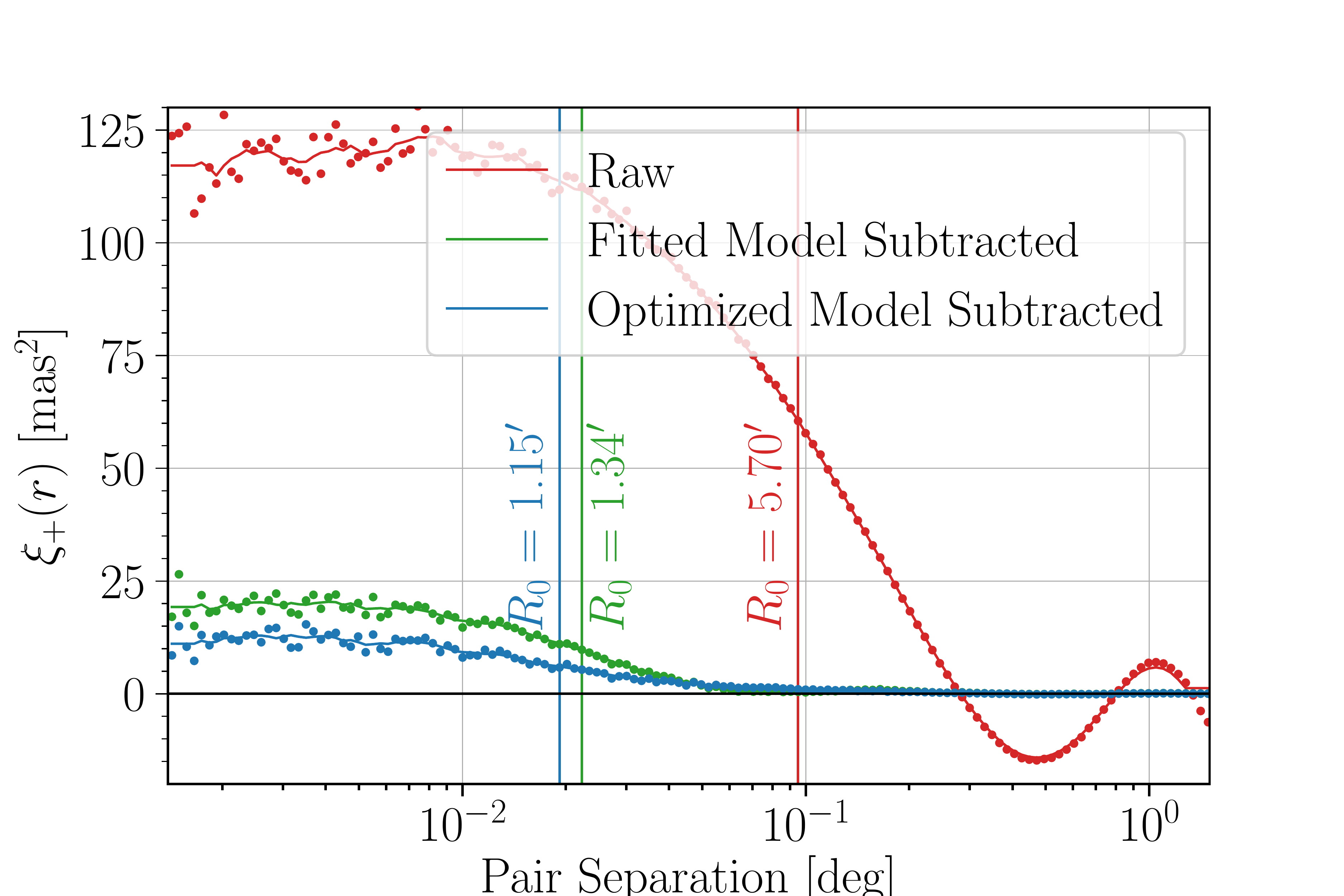}
  \caption{The points plot the mean azimuthally-averaged $\xi_+(\lvert
    \vx \rvert)$ for 206
    $riz$-band exposures.  The curves apply a Savitsky-Golay smoothing
    to the (noisy) measured points.  
    separation from $5\arcsec$ to 1.5~deg. From top to bottom, the
    curves show the correlated astrometric errors before any GPR
    subtraction (``Raw''), after subtracting a GPR using the best-fit
    von Karman turbulence parameters (``Fitted model''), and after
    subtracting a GPR using an ``optimized'' kernel chosen to minimize
    $\xi_0=\xi_+(\vx\rightarrow 0).$ The astrometric variance is greatly
    reduced from the raw level of $\xi_0=119$~mas$^2$ to $19.5$ and
    $12.0$~mas$^2$ by the fitted and optimized kernels, respectively.
    The average correlation length of the astrometric errors (defined as the
    point where $\xi_+$ drops to $\xi_0/2$) is reduced from 5\farcm7
    to $1\farcm2$ by GPR subtraction.}
  \label{fig:avgxi}
\end{figure}

Our primary measure of success for this analysis is the reduction of the $\xi_0=\langle u^2 + v^2\rangle$ of the raw residual field after the GPR model of the turbulence is subtracted. With a finite number of validation stars, we cannot calculate $\xi_0(\vx=0)$.
To measure the improvements due to GPR subtraction, we
approximate $\xi_0$ with $\xi_+(\vert \vx \rvert < 0.5\arcmin)$, \ie\
we average over all validation-star pairs with separation
$<0.5\arcmin.$\footnote{Note that this $\xi_0$ estimator uses a
  smaller radius (0\farcm5) than the $\xi_{1.2}$ used during kernel
  optimization.  This is because Figure~\ref{fig:avgxi} shows that
  $\xi(\vx)$ is already dropping significantly at $|\vx|=1\farcm2$ in
  the GPR-subtracted data.}
Figure~\ref{fig:xiBA} plots the raw $\xi_0$ vs the GPR-subtracted
$\xi_0$ for $rizY$ exposures, with the lower panel replotting the
quantity $\sqrt{\xi_0/2},$ which gives the RMS astrometric error per
axis. The lower panel also shows the relationship between RMS and field density for $rizY$ exposures. The average $\xi_0$ of the raw residual field for all 206 $riz$-band exposures was 125 mas$^2$, or an RMS astrometric error of 7.3 mas per axis. The average $\xi_0$ after the GPR model was subtracted from the residual field was 11.9 mas$^2$ or 2.3 mas RMS. The average reduction in $\xi_0$ was a factor of 12. This is the principal result of this work.

We also note that the kernel optimization of Step~\ref{item:optimization} reduces by $\xi_0$ by a mean factor of $\approx1.6\times$ for $riz$-band exposures (with a wide range of variation), and it is possible that this optimization could be improved.

Figure~\ref{fig:avgxi} plots $\xi_+(|\vx|)$ averaged over all $riz$
exposures, showing that (per expectations) the longer-range
correlations are completely eliminated by GPR subtraction.
The correlation length, $R_0$, is defined as the angle such that $\xi_+(R_0) = 0.5\xi_0$.
For the 206 $riz$-band exposures, the average correlation length of the raw residual field was $5\farcm7$. The average correlation length of the residual field after the GPR model was subtracted was $1\farcm3$. As expected, the post-GPR correlation length is similar to the mean distance between \gaia\ stars used for training.

The average von-Karman kernel parameters for the $riz$ functions are in rough agreement with a physical model, although it is clear from \eg\ Figure~\ref{fig:xi2d} that the single-screen turbulence model is an incomplete description of the astrometric correlation functions.  The average  aperture-diameter parameter, $d$, was $2.5\arcmin,$ which corresponds to the physical 4-meter diameter $D$ for a turbulence height of 5.5~km, which is of similar magnitude to the typical height of dominant turbulence.  At this height the average angular outer-scale parameter of $r_0=1\fdg7$ corresponds to 160~m, also physically reasonable---though this parameter has weak effect on the kernel or GPR accuracy.  The typical wind parameter amplitude of $|\boldsymbol{w}|=3\farcm5$ corresponds to winds of just $\approx0.1$~m~s$^{-1},$ which seems lower that one would expect for tropospheric winds.


There are few clear trends in the values of the post-GPR $\xi_0$ or in
the reduction factor gained by the GPR.  From the lower panel Figure~\ref{fig:xiBA} it is clear that the exposures taken in low-density regions of the \gaia\ catalog have the lowest (worst) reduction factors.  This is to be expected, as density \gaia\ training data can measure and remove the turbulence to higher spatial frequencies.  We also find, not surprisingly, that a lower raw $\xi_0$ correlates with a lower GPR-subtracted $\xi_0.$  Beyond this, there are no obvious trends with filter band or other variables.
The sizes of $\xi_0$ both before and after GPR subtraction are highly variable over time, as one finds for related atmospheric-turbulence phenomena such as the seeing FWHM.  These large ``weather'' variations could be masking subtler trends with wavelength, airmass, etc.\ that might emerge upon analysis of a larger number of exposures.

\begin{figure}[ht]
  \centering
\end{figure}

\subsection{Test on Eris orbit}
As a test of the GPR astrometric correction, we examine the residuals
of an orbit fit to the positions measured by \des\ for the
trans-Neptunian object Eris \citet{discoverEris}.  Because Eris moves $\approx1\fdg4$ in
both RA and Dec over the 5-year span of the \des\ observations, it is
being informed by a continuously changing set of \gaia\ stars, and
thus the orbit residuals are a good sampling of the accuracy of the
GPR. A good orbit also requires a correct absolute astrometric
calibration, at least across these few degrees.

\begin{figure*}[ht]
  \includegraphics[width=\textwidth]{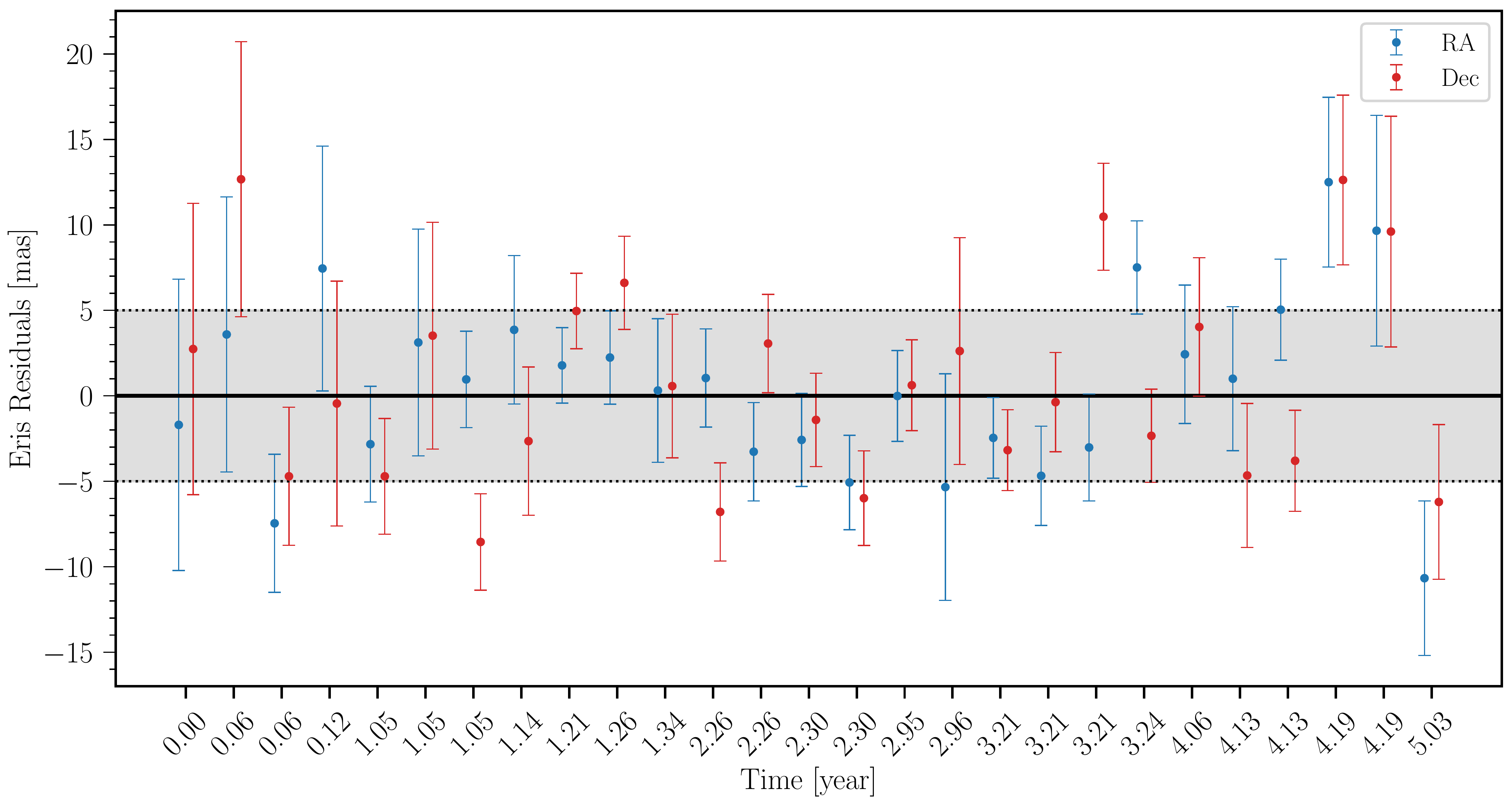}
  \caption{Residuals of the observed RA/Dec positions from
    GPR-corrected \des\ measurements of the trans-Neptunian object
    Eris to the best-fitting orbit are plotted.  The observations span
    5 years, with the relative times marked below each point pair.
    The RMS error of the $riz$-band observations is 5~mas, as marked
    by the shaded region, consistent with the estimated shot noise
    plus residual turbulence $\xi_0$.  The RMS residual was 10~mas
    before the GPR correction was made, demonstrating the success of
    the GPR method.}
  \label{fig:eris}
\end{figure*}

Figure~\ref{fig:eris} plots the residuals to the 27 $griz$ observations of Eris in the \des\ Wide survey, compared to the best-fit orbit as obtained using the algorithms of \citet{bk}.  The error bars in the plot are the expected uncertainties from the quadrature sum of the \texttt{ERRAWIN\_WORLD} measurement error and the $\xi_0$-derived RMS turbulence error.  The measured pixel positions for the Eris detections are converted to RA and Dec using the full static astrometric model as described above, including chromatic terms.

Restricting our consideration to the 19 $riz$ exposures, the RMS
residual to the best-fit orbit is 10.1~mas (per axis) using the raw
positions (no turbulence correction), but drops to 5.0~mas after the
GPR estimates are subtracted.  The $\chi^2$ values and degrees of
freedom are 27.9 and 44.7/32 before and after GPR subtraction, respectively.  This demonstrates a clear improvement in astrometric quality.  The factor of 2 improvement in RMS is less than the $\sqrt{12}$ we might expect from the typical reduction in $\xi_0,$ because Eris, at $r=18.5$~mag, is not bright enough to be in the fully turbulence-dominated regime for \des\ once the GPR is applied: the shot-noise errors in its positions vary from 1.7--5.2~mas (with an RMS value of 2.9~mas).  Indeed, once the GPR correction is applied, a star must be near the exposure saturation limit in order to have shot noise well below the turbulence noise.

These measurement accuracies are far better than is typically obtained for minor planets from ground-based observations, especially considering the short (90 s) exposure times.  To give a sense of scale, a typical trans-Neptunian object at 40~AU distance moves by 4~mas per second, so shutter-timing corrections are now larger than the $\approx2$~mas astrometric turbulence error.

\section{Summary and prospects}
\label{sec:conclusion}

We have demonstrated that Gaussian processes are highly effective for
interpolating the stochastic astrometric distortions from a set of
known spatial points (\gaia\ stars) to arbitrary locations in the
focal plane.  For \des\ images in $riz$ bands, we achieve an average
reduction of a factor $\approx12$ in the total astrometric variance
$\xi_0$ ascribable to turbulence.  As expected, the GPR is very
successful at modelling (and removing) distortion modes at wavelengths
longer than the typical $\approx1\arcmin$ spacing between \gaia\ DR2
stars, and the GPR-subtracted astrometric errors have a correlation length of
$\approx1\arcmin.$  This achievement is assisted by a new variant of
GPR which makes use of the known curl-free nature of the 2d distortion
field. 

For these 90~s images, this reduces
the RMS turbulence error to $\approx2$~mas in each axis, at which
point it is subdominant to shot noise in the object centroid for any
source with $S/N\lesssim 200$, which requires $>40,000$ photoelectrons
to be acquired.  There is now thus only a $\approx1$~mag range of
stellar brightness in which the astrometry is dominated by turbulence
noise but the stars are not saturating the CCD, so in practical terms
the atmospheric turbulence noise has been nearly eliminated.

It is likely that some further improvement is
possible by improving the optimization of the von Karman parameters,
or by choosing a better form for the kernel. It is already clear from
Figure~\ref{fig:xi2d} that the observed $\xi_+(\vx)$ function has
qualitative aspects not reproduced by the von Karman model, such as
different major axes at different scales, which could result from
multiple layers of turbulence.  Such behavior has been observed in
data from the Canada-France-Hawaii Telescope, and it is worth
investigating whether using the measured $\xi_+$ directly, rather than
a model fit, is generically better (P. F. Leget, private
communication).  We have not developed this yet, because of the
possibility that a noisy measured $\xi_+$ could lead to
non-positive-definite covariance matrices.

Dramatic improvement may be possible by
creating a denser training set than \gaia\ DR2 provides.  Future releases
of \gaia\ will help in this regard.  But a more powerful means to
bootstrap a denser training sample
is outlined in the Appendix: when we require that non-\gaia\ stars'
motion over time fits the standard parallax/proper-motion model, their
5d solutions become increasing constrained with more observing epochs,
and the exposure-by-exposure residuals to these 5d solutions become
useful for constraining the turbulence patterns of individual
exposures.  The Appendix gives the mathematical solution for a
feasible scheme to solve for the 5d solutions and the turbulence
interpolation for all exposures in a region of sky simultaneously.  We
will test this in future work.  The astrometric system, \ie\ the
absolute coordinates, proper-motion, and parallax zeropoints, will remain
tied to \gaia, and here again ground-based results will benefit from
future \gaia\ releases.

With the joint turbulence/proper-motion solution, essentially
  every star with per-exposure shot noise level lower than the
  turbulence noise becomes an additional reference point for the
  turbulence solution.  While DES has few stars fainter than the
  \gaia\ limit which satisfy this condition, \lsst\
  will have many.  We thus expect this to be a valuable technique for the Survey.

Note that both shot-noise errors and turbulence residuals are expected
to decrease as the inverse square root of exposure time $T.$  Thus for
a given observatory and filter band, the magnitude dividing
turbulence-dominated from shot-noise-dominated astrometry is
independent of $T.$  The GPR technique shifts that magnitude brightward by
about 2.5~mag.  For longer DECam exposures, both limits drop, and
astrometry could become limited by sub-mas errors in the static astrometric
model, including errors in chromatic corrections arising from
inaccurate knowledge of the source spectra.

This GPR technique should be directly applicable to the \lsst\ data.
The Rubin Observatory's 8.5-meter-diameter primary mirror is larger than the
Blanco's 4-meter primary, which will lower the expected level of
$\xi_+$ in similar atmospheric conditions; but the shorter total exposure time
(30~s) of nominal \lsst\ visits will increase the per-exposure $\xi_+,$
so we might expect values similar to those we find for \des.  Rubin's
greater aperture means, however, that shot noise is lower than for
\des, which means that more stars---including many beyond \gaia's
magnitude limit---will be in the turbulence-dominated regime and thus
benefit from the GPR subtraction.  This should allow \lsst\ data to
substantially surpass the initial requirements and goals for its
astrometric performance and subsequent science.

\acknowledgments

We thank Pierre-Francois Leget for very useful conversations about
anisotropic GP's and 
atmospheric turbulence at other telescopes.  University of
Pennsylvania authors have been supported in this work by grants
AST-1515804, AST-1615555, and AST-2009210 from the National Science Foundation, and
grant DE-SC0007901 from the Department of Energy. 

Funding for the DES Projects has been provided by the U.S. Department of Energy, the U.S. National Science Foundation, the Ministry of Science and Education of Spain, 
the Science and Technology Facilities Council of the United Kingdom, the Higher Education Funding Council for England, the National Center for Supercomputing 
Applications at the University of Illinois at Urbana-Champaign, the Kavli Institute of Cosmological Physics at the University of Chicago, 
the Center for Cosmology and Astro-Particle Physics at the Ohio State University,
the Mitchell Institute for Fundamental Physics and Astronomy at Texas A\&M University, Financiadora de Estudos e Projetos, 
Funda{\c c}{\~a}o Carlos Chagas Filho de Amparo {\`a} Pesquisa do Estado do Rio de Janeiro, Conselho Nacional de Desenvolvimento Cient{\'i}fico e Tecnol{\'o}gico and 
the Minist{\'e}rio da Ci{\^e}ncia, Tecnologia e Inova{\c c}{\~a}o, the Deutsche Forschungsgemeinschaft and the Collaborating Institutions in the Dark Energy Survey. 

The Collaborating Institutions are Argonne National Laboratory, the University of California at Santa Cruz, the University of Cambridge, Centro de Investigaciones Energ{\'e}ticas, 
Medioambientales y Tecnol{\'o}gicas-Madrid, the University of Chicago, University College London, the DES-Brazil Consortium, the University of Edinburgh, 
the Eidgen{\"o}ssische Technische Hochschule (ETH) Z{\"u}rich, 
Fermi National Accelerator Laboratory, the University of Illinois at Urbana-Champaign, the Institut de Ci{\`e}ncies de l'Espai (IEEC/CSIC), 
the Institut de F{\'i}sica d'Altes Energies, Lawrence Berkeley National Laboratory, the Ludwig-Maximilians Universit{\"a}t M{\"u}nchen and the associated Excellence Cluster Universe, 
the University of Michigan, NFS's NOIRLab, the University of Nottingham, The Ohio State University, the University of Pennsylvania, the University of Portsmouth, 
SLAC National Accelerator Laboratory, Stanford University, the University of Sussex, Texas A\&M University, and the OzDES Membership Consortium.

Based in part on observations at Cerro Tololo Inter-American Observatory at NSF’s NOIRLab (NOIRLab Prop. ID 2012B-0001; PI: J. Frieman), which is managed by the Association of Universities for Research in Astronomy (AURA) under a cooperative agreement with the National Science Foundation.

The DES data management system is supported by the National Science Foundation under Grant Numbers AST-1138766 and AST-1536171.
The DES participants from Spanish institutions are partially supported by MICINN under grants ESP2017-89838, PGC2018-094773, PGC2018-102021, SEV-2016-0588, SEV-2016-0597, and MDM-2015-0509, some of which include ERDF funds from the European Union. IFAE is partially funded by the CERCA program of the Generalitat de Catalunya.
Research leading to these results has received funding from the European Research
Council under the European Union's Seventh Framework Program (FP7/2007-2013) including ERC grant agreements 240672, 291329, and 306478.
We  acknowledge support from the Brazilian Instituto Nacional de Ci\^encia
e Tecnologia (INCT) do e-Universo (CNPq grant 465376/2014-2).

This manuscript has been authored by Fermi Research Alliance, LLC
under Contract No. DE-AC02-07CH11359 with the U.S. Department of
Energy, Office of Science, Office of High Energy Physics.

\emph{Software}: The software developed in this work will be made public shortly after the publication. This work made use of the following public codes: \textsc{Numpy} \citep{Numpy}, \textsc{SciPy} \citep{SciPy}, \textsc{Astropy} \citep{,Astropy18}, \textsc{Matplotlib} \citep{Matplotlib}, \textsc{IPython} \citep{iPython}, \textsc{TreeCorr} \citep{Treecorr}, \textsc{easyaccess} \citep{Easyaccess}, \textsc{WCSFit} and \textsc{pixmappy} \citep{decamast}, 

\bibliography{references}
\bibliographystyle{aasjournal}

\appendix
\section{Simultaneous inference of turbulence and 5d stellar
  parameters}
\label{sec:bigsolution}
The positions of isolated stars on the sky are expected to follow the standard
5-parameter model:
\begin{align}
  \vx^{\rm true}_{i\mu}  & =  \vx^{0}_{i}  + t_\mu \dot\vx_i -
                           \vx^E_{\perp\mu} \varpi_i. \\
   & = \matM_\mu \vs_i.
\end{align}
Here $t_\mu$ and $\vx^E_{\perp\mu}$ are the date of exposure $\mu$,
and the projection of the barycentric observatory position onto the
line of sight for the exposure.
The 5 parameters for star $i$ are the position, proper
motion, and parallax $\vs_i = \{x^0_i, y^0_i, \dot x_i, \dot y_i,
\varpi_i\}.$ 
The goal of this section is to develop a method for extracting $\vs =
\{\vs_1,\vs_2,\ldots\}$ from the observed positions $\vx^{\rm
  obs}_{i\mu}$ of each star in each exposure.\footnote{A given star
  need not be observed in all exposures.}  The data model is
\begin{equation}
  \vx^{\rm obs}_{i\mu} = \matM_\mu \vs_i + \vu_{i\mu}.
\label{pm5}
\end{equation}
We will always assume that $\vx^{\rm obs}$ has been mapped from pixel
coordinates to sky
coordinates using the best available static instrument model, and \vu\
is a stochastic, zero-mean error term that includes curl-free atmospheric
turbulence, shot noise, and any errors in the instrumental model.

The method of Section~\ref{curlfree} gives a straightforward procedure
by which a GPR trained on the \gaia\ star images in exposure $\mu$
yields an estimator $\hat\vu_{i\mu}$ for each individual image of a non-\gaia\
star.  One could then fit the set of observed $ \vx^{\rm
  obs}_{i\mu}-\hat\vu_{i\mu}$ to the model (\ref{pm5}) to estimate
$\vs_i.$
It is the case, however, that as more exposures are taken and the
$\vs_i$ becomes better known, that this star can begin to inform the
training of the GPR for each exposure, improving the estimator
$\hat\vu$ for other stars in its vicinity, and in turn improving their
5d solutions.  We therefore explore the possibility of an estimator
for \vs\ that considers the data from all exposures simultaneously.

To do so, we continue the assumption that the displacement field $\vu_\mu$ is
the sum of a turbulence contribution, which is the gradient of a
Gaussian random field with known power spectrum, and a shot noise
contribution, which does not correlate between stellar observations.
We further assume that there is no correlation of either component
between distinct exposures---and each exposure can have its own
correlation function $K_\mu,$ determined by hyperparameter optimization on the
\gaia\ stars, as described above.

In this case the probability of the observations is a multivariate
Gaussian that separates between exposures:
\begin{equation}
  p\left(\{\vx^{\rm obs}_\mu\} | \vs \right) = \prod_\mu \left|2\pi \matK_\mu
  \right|^{-1/2}   \exp\left[ -\frac{1}{2} \left(\vx^{\rm obs}_\mu - \matM_\mu \vs\right)^T
    \matK_\mu^{-1}  \left(\vx^{\rm obs}_\mu - \matM_\mu \vs\right)\right]
\label{pobs2}
\end{equation}
where we have used the data model in \eqq{pm5}, and defined
\begin{equation}
  K_{\mu,ij} = K_{\mu}(\vx_{i\mu}-\vx_{j\mu}) + n_{i\mu} \delta_{ij}
\end{equation}
Also, $\vx^{\rm obs}_\mu$ is the concatenation of
all the observed positions on exposure $\mu$.

Bayes' theorem, assuming independent Gaussian priors on each star's
parameters, gives
\begin{equation}
  p\left(\vs | \{\vx^{\rm obs}_\mu\}\right) \propto
  p\left(\{\vx^{\rm obs}_\mu\} | \vs \right)  \times \prod_i {\cal N}( \vs^p_i,
  \matK^p_i).
\label{bayes2}
\end{equation}
The prior can include \gaia\ measurements of the star's $\vs_i,$ if available. It is also
advisable to place a weak prior on the parallax for non-\gaia\ stars
to avoid numerical instabilities for stars that have been observed at
the same limited range of dates each year.
The observing cadence of \des\ is not designed
for parallax measurements and can produce such degeneracies.

By substituting \eqq{pobs2} into this equation, and concatenating the
stellar parameters and their
priors into $\vs^p = \{\vs_1,\vs_2,\ldots\},$ and $\matK^p =
\mathbf{diag}\left(\{\matK^p_i\}\right),$ we obtain the log posterior
for the stellar parameters:
\begin{align}
  -2\log p\left(\vs | \{\vx^{\rm obs}_\mu\},\vs^p,\matK_p\right) & = 
(\textrm{const}) + \left(\vs - \vs^p\right)^T \matK_p^{-1} \left(\vs -
                                                     \vs^p\right) \\
   & \phantom{=} + 
\left(\vx^{\rm obs}_\mu - \matM_\mu \vs\right)^T
     \matK_\mu^{-1}  \left(\vx^{\rm obs}_\mu - \matM_\mu \vs\right) \\
  \Rightarrow \qquad \vs | \{\vx^{\rm obs}_\mu\},\vs^p,\matK_p & \sim {\cal N}\left(
                                                   \hat\vs,
                                                   \matC_s\right), \\
  \matC_s & = \left[ \matK_p^{-1} + \sum_\mu \matM^T_\mu
            \matK_\mu^{-1} \matM_\mu \right]^{-1}  \label{covs}\\
  \hat\vs & = \matC_s \left[ \matK_p^{-1} \vs^p +  \sum_\mu \matM^T_\mu
            \matK_\mu \vx^{\rm obs}_\mu \right].
            \label{gpr2}
\end{align}
The last line gives the maximum-posterior estimate of all the stars'
5d properties.  This joint solution is computationally feasible; GPR's
are rate-limited by Cholesky matrix factorizations.
If there are $N_s$ stars appearing in $N_e$ exposures, then each of
the $\matK_\mu$ are $2N_s \times 2N_s$ square, so the sum inside the
brackets of (\ref{covs}) requires
$O(8N_eN_s^3)$ operations. 
The $\hat\vs$ vector has $5N_s$ elements,
so the factorization needed to obtain it in (\ref{gpr2}) requires $O(125N_s^3)$
operations, which in most cases ($N_e>15$) is lower than the cost of
the per-exposure inversions.

It is clearly infeasible to execute the solution in one
shot over the full DES survey, with $N_s\sim 10^8$ and $N_e\sim10^5.$
There is little lost, however, in dividing the survey into regions
and solving each independently.  Each region's solution is informed
by stars within a few atmospheric correlation lengths of its targeted
boundaries, which should be included in the solution.
But more distant stellar measures do not improve the
target region and need not be included.
Thus the computation is best managed by dividing the
survey into $\approx 1\deg$ regions, with $N_s~10^5, N_e\sim10^2.$

The computational cost of the wholistic solution can be compared to
the cost of optimizing the hyperparameters of $\matK_\mu$ for each
exposure.  The optimizer must re-invert the $(2\times N_G)^2$ matrix
many times to maximize the log marginal likelihood, with $N_G$ being
the number of \gaia\ stars in the full exposure, whereas the
wholistic solution requires a single $(2\times N_s)^2$ inversion,
where $N_s$ is the total number of stars of interest in the
region.  In most cases wholistic solution will take comparable or less
time than completing the per-exposure kernel tuning.

The only distinction between \gaia\ detections and
\des-only stars in \eqq{gpr2} is that the former have stronger priors.  The
solution uses all of the stars to constrain the turbulence field.
In practice one will also want to interpolate the turbulence field to
the locations of sources in individual exposures which do not have 5d
solutions (\eg\ minor planets and other transients).  As with a simple
GPR, it is straightforward to perform this interpolation, and the
computational cost is much lower than for the Cholesky inversion of
$\matC_s.$

\allauthors
\end{document}